\documentclass[zpreprint,zbstepj]{./LaTeX/zeus/zeus_paper}
\usepackage[english]{babel}
\usepackage{axodraw}
\usepackage[T1]{fontenc}
\newcommand{\ZcoosysA}{%
The ZEUS coordinate system is a right-handed Cartesian system, with the $Z$
axis pointing in the proton beam direction, referred to as the ``forward
direction'', and the $X$ axis pointing left towards the centre of HERA.
The coordinate origin is at the nominal interaction point.\xspace}

\chardef\usc=95
\chardef\til=126
\catcode`\@=11 
\DeclareRobustCommand\xdotspace{\futurelet\@let@token\@xdotspace}
\def\@xdotspace{%
  \ifx\@let@token.\else
  \ifx\@let@token\bgroup.\else
  \ifx\@let@token\egroup.\else
  \ifx\@let@token\/.\else
  \ifx\@let@token\ .\else
  \ifx\@let@token~.\else
  \ifx\@let@token!.\else
  \ifx\@let@token,.\else
  \ifx\@let@token:.\else
  \ifx\@let@token;.\else
  \ifx\@let@token?.\else
  \ifx\@let@token/.\else
  \ifx\@let@token'.\else
  \ifx\@let@token).\else
  \ifx\@let@token-.\else
  \ifx\@let@token\@xobeysp.\else
  \ifx\@let@token\space.\else
  \ifx\@let@token\@sptoken.\else
   .\space
   \fi\fi\fi\fi\fi\fi\fi\fi\fi\fi\fi\fi\fi\fi\fi\fi\fi\fi}
\catcode`\@=12 

\newcommand{\stru}[2]{%
   \relax\ifmmode\hbox{\vrule height#1 depth#2 width0pt}%
   \else\vrule height#1 depth#2 width0pt\fi}

\newcommand{\Ronum}[1]{\uppercase\expandafter{\romannumeral#1}}
\newcommand{\ronum}[1]{\expandafter{\romannumeral#1}}
\DeclareRobustCommand{\LaTeXZ}{%
  \LaTeX\kern-.05em4\kern-.1em
  {\raisebox{-0.2ex}{$\scriptstyle\text{ZEUS}$}}\xspace}

\DeclareMathAlphabet{\mathbf}{OT1}{cmr}{bx}{sl}
\newcommand{\eVdist}{\kern-0.06667em}

\newcommand{\Gev}{{\text{Ge}\eVdist\text{V\/}}}

\newcommand{\slashfrac}[2]{%
  \raisebox{0.5ex}{\ensuremath #1}\kern-0.12em/\kern-0.08em
  \raisebox{-.8ex}{\ensuremath #2}}

\newcommand{\sqr}[3]{%
    {\vcenter{\hrule height.#3ex\hbox{\vrule width.#2ex height#1ex
     \kern#1ex\vrule width.#3ex}\hrule height.#2ex}}}

\catcode`\@=11 
\newcommand{\parenbar}{\mathpalette\p@renb@r}
\def\p@renb@r#1#2{\vbox{%
  \ifx#1\scriptscriptstyle \dimen@.7em\dimen@ii.2em\else
  \ifx#1\scriptstyle \dimen@.8em\dimen@ii.25em\else
  \dimen@1em\dimen@ii.4em\fi\fi \offinterlineskip
  \ialign{\hfill##\hfill\cr
    \vbox{\hrule width\dimen@ii}\cr
    \noalign{\vskip-.3ex}%
    \hbox to\dimen@{$\mathchar300\hfil\mathchar301$}\cr
    \noalign{\vskip-.3ex}%
    $#1#2$\cr}}}
\catcode`\@=12 

\newcommand{\diff}{{\rm d}}

\newcommand{\Born}{{\rm Born}}

\newcommand{\IP}{{\rm I$\kern-0.01667em$P}\xspace}

\mathchardef\qsm=63
\mathchardef\pls=43
\mathchardef\mns=512
\mathchardef\plm=518
\mathchardef\eql=61
\mathchardef\smallleft=300
\mathchardef\smallright=301
\mathchardef\les=316
\mathchardef\gre=318
\mathchardef\leq=532
\mathchardef\grq=533

\catcode`\@=11 
\newcounter{pict@width}
\newcounter{pict@height}
\newlength{\pict@scale}
\newcommand{\psfigadd}[4]{%
\setcounter{pict@width}{1*\ratio{#2+\pict@scale/2}{\pict@scale}}
\setcounter{pict@height}{1*\ratio{#3+\pict@scale/2}{\pict@scale}}
\setlength{\unitlength}{\pict@scale}
\hbox to #2{\hspace{-\fill}\begin{picture}(\thepict@width,\thepict@height)
\put(0,0){\psfig{figure=#1,width=#2,height=#3,clip=}}
\SetScale{0.283466457}
\SetWidth{1.763889}
{#4}
\end{picture}}
}
\newcounter{pict@widthfst}
\newcounter{pict@widthscd}
\newcounter{pict@widthtot}
\newcommand{\psfigaddtwo}[7]{%
\setcounter{pict@widthfst}{1*\ratio{#2+\pict@scale/2}{\pict@scale}}
\setcounter{pict@widthscd}{1*\ratio{#2+#4+\pict@scale/2}{\pict@scale}}
\setcounter{pict@widthtot}{1*\ratio{#2+#4+#6+\pict@scale/2}{\pict@scale}}
\setcounter{pict@height}{1*\ratio{#3+\pict@scale/2}{\pict@scale}}
\setlength{\unitlength}{\pict@scale}
\hbox{\hspace{-\fill}\begin{picture}(\thepict@widthtot,\thepict@height)
\put(0,0){\psfig{figure=#1,width=#2,height=#3,clip=}}
\put(\thepict@widthscd,0){\psfig{figure=#5,width=#6,height=#3,clip=}}
\SetScale{0.283466457}
\SetWidth{1.763889}
{#7}
\end{picture}}
}
\newcommand{\psfigror}[4]{%
\setcounter{pict@width}{1*\ratio{#2+\pict@scale/2}{\pict@scale}}
\setcounter{pict@height}{1*\ratio{#3+\pict@scale/2}{\pict@scale}}
\setlength{\unitlength}{\pict@scale}
\hbox{\begin{picture}(\thepict@width,\thepict@height)
\put(0,\thepict@height){\psfig{figure=#1,width=#3,height=#2,clip=,angle=270}}
\SetScale{0.283466457}
\SetWidth{1.763889}
{#4}
\end{picture}}
}
\newcommand{\psfigrol}[4]{%
\setcounter{pict@width}{1*\ratio{#2+\pict@scale/2}{\pict@scale}}
\setcounter{pict@height}{1*\ratio{#3+\pict@scale/2}{\pict@scale}}
\setlength{\unitlength}{\pict@scale}
\hbox{\begin{picture}(\thepict@width,\thepict@height)
\put(0,0){\psfig{figure=#1,width=#3,height=#2,clip=,angle=90}}
\SetScale{0.283466457}
\SetWidth{1.763889}
{#4}
\end{picture}}
}
\catcode`\@=12 
\newlength\listtextwidth

\catcode`\@=11 
\newlength{\@tabfninsert}
\newlength{\@tabfnwidth}
\newcommand{\tabfootnote}[2]{%
  \setlength{\@tabfninsert}{0.8em}
  \setlength{\@tabfnwidth}{\textwidth}
  \addtolength{\@tabfnwidth}{-\@tabfninsert}
  \addtolength{\@tabfnwidth}{-0.4em}
  \noindent\makebox[\@tabfninsert][r]{\footnotesize$^{#1}$\hfil}\hfill%
  \parbox[t]{\@tabfnwidth}{\footnotesize #2\hfill}}
\catcode`\@=12 

\newcommand{\plot}[4]{
  \begin{figure}[p]
  \vfill
  \begin{center}
    \begin{picture}(450,550)(0,0)
    \put(0,#2){\mbox{\epsfig{file=#1,width=450pt}}}
    \end{picture}
  \end{center}
      \caption[]{#4}
  \vfill
  \label{#3}
  \end{figure}
}
\newcommand{\plotup}[3]{
  \begin{figure}[p]
  \vfill
  \begin{center}
    \begin{picture}(430,500)(0,0)
    \put(0,-20){\mbox{\epsfig{file=#1,width=430pt}}}
    \PText(65,420)(0)[c]{(a)}
    \PText(65,150)(0)[c]{(b)}
    \end{picture}
  \end{center}
      \caption[]{#3}
  \vfill
  \label{#2}
  \end{figure}
}

\newcommand{\ppth}      {\mbox{$P^2_{T,h}$}}
\newcommand{\pth}       {\mbox{$P_{T,h}$}}
\newcommand{\pxh}       {\mbox{$P_{X,h}$}}
\newcommand{\pyh}       {\mbox{$P_{Y,h}$}}
\newcommand{\pzh}       {\mbox{$P_{Z,h}$}}
\newcommand{\ppte}       {\mbox{$P^2_{T,e}$}}
\newcommand{\pte}       {\mbox{$P_{T,e}$}}

\newcommand{\ypt}       {\mbox{$y_{\textrm{PT}}$}}
\newcommand{\xpt}       {\mbox{$x_{\textrm{PT}}$}}
\newcommand{\qpt}       {\mbox{$Q^2_{\textrm{PT}}$}}
\newcommand{\wpt}       {\mbox{$W_{\textrm{PT}}$}}
\newcommand{\gpt}       {\mbox{$\gamma_{\textrm{PT}}$}}
\newcommand{\zvtx}      {\mbox{$Z_{\textrm{vertex}}$}}
\newcommand{\FL}        {\mbox{$F_L$}}
\newcommand{\perc}      {~\mbox{\hspace*{-0.3em}\%}}
\newcommand{\Gevsq}     {\mbox{${\rm GeV}^2$}}
\newcommand{\Fem}       {\mbox{$F_2^{em}$}}
\newcommand{\Fint}      {\mbox{$F_2^{int}$}}
\newcommand{\Fwk}       {\mbox{$F_2^{wk}$}}
\newcommand{\Ft}        {\mbox{$F_2$}}
\newcommand{\Fd}        {\mbox{$xF_3$}}

\newcommand{\qsd}       {\mbox{${Q^2}$}}
\newcommand{\x}         {\mbox{${\it x}$}}
\newcommand{\y}         {\mbox{${\it y}$}}
\newcommand{\ye}        {\mbox{${y_{e}}$}}

\newcommand{\yjb}       {\mbox{${y_{_{JB}}}$}}

\newcommand{\gammah}    {\mbox{$\gamma_{_{h}}$}}

\newcommand{\gsim}      {\mbox{\raisebox{-0.4ex}{$\;\stackrel{>}{\scriptstyle \sim}\;$}}}
\newcommand{\lsim}      {\mbox{\raisebox{-0.4ex}{$\;\stackrel{<}{\scriptstyle \sim}\;$}}}

\def\3{\ss}

\def\citeCTD{{\cite{%
nim:a279:290,*npps:b32:181,*nim:a338:254%
}}\xspace}
\def\citeCAL{{\cite{%
nim:a309:77,*nim:a309:101,*nim:a321:356,*nim:a336:23%
}}\xspace}

\def\citeLUMI{{\cite{%
desy-92-066,*zfp:c63:391%
}}\xspace}

\def\citeCFLT{{\cite{%
nim:a355:278%
}}\xspace}

\def\citeZEUSF294{{\cite{%
zfp:c72:399%
}}\xspace}
\def\citeH1F2{{\cite{%
np:b470:3%
}}\xspace}
\def\citeZEUSF2lowx{{\cite{%
pl:b407:432%
}}\xspace}
\def\citeH1F2lowx{{\cite{%
np:b497:3%
}}\xspace}
\def\citeZEUShighq2{{\cite{%
epj:c11:427%
}}\xspace}
\def\citeH1highq2{{\cite{%
epj:c13:609%
}}\xspace}

\def\citeCDM{{\cite{%
pl:b165:147,*np:b306:746,*zfp:c43:625%
}}\xspace}

\begin{document}                                                 
\title{Measurement of the neutral current cross section 
and $\mathbf{F_2}$ structure function
for deep inelastic $\mathbf{e^+p}$ scattering
at HERA\\
\vspace{-6.5cm}
\begin{flushleft}
\tt\normalsize DESY 01-064\\
May\ 2001\\[5cm]
\end{flushleft}
}

\author{ZEUS Collaboration}
\draftversion{12.0}
\date{June 19th, 2001}

\abstract{
    The cross section
    and the proton structure function $F_2$ 
    for neutral current deep inelastic $e^+p$ scattering
    have been measured
    with the ZEUS detector at HERA 
    using an integrated luminosity of 30 pb$^{-1}$.
    The data
    were collected in 1996 and 1997
    at a centre-of-mass energy of 300 GeV.
    They cover the kinematic range
    $2.7 < Q^2 < 30000$ GeV$^2$
    and $6\cdot 10^{-5} < x < 0.65$.
    The variation of 
    $F_2$ with $x$ and $Q^2$
    is  well described by next-to-leading-order 
    perturbative QCD as implemented in
    the DGLAP evolution equations.
}

\makezeustitle

\newcommand{\address}{ }                                                                           
\newcommand{\author}{ }                                                                          
\pagenumbering{Roman}                                                                              
\begin{center}                                                                                     
{                      \Large  The ZEUS Collaboration              }                               
\end{center}                                                                                       
  S.~Chekanov,                                                                                     
  M.~Derrick,                                                                                      
  D.~Krakauer,                                                                                     
  S.~Magill,                                                                                       
  B.~Musgrave,                                                                                     
  A.~Pellegrino,                                                                                   
  J.~Repond,                                                                                       
  R.~Stanek,                                                                                       
  R.~Yoshida\\                                                                                     
 {\it Argonne National Laboratory, Argonne, IL, USA}~$^{p}$                                        
\par \filbreak                                                                                     
  M.C.K.~Mattingly \\                                                                              
 {\it Andrews University, Berrien Springs, MI, USA}
% ...
\newpage                                                                                           
$^{\    1}$ now visiting scientist at DESY \\                                                      
$^{\    2}$ now at Univ. of Salerno and INFN Napoli, Italy \\                                      
% ...
                                                           %
                                                           %
\newpage   
                                                           %
                                                           %
\begin{tabular}[h]{rp{14cm}}                                                                       
$^{a}$ &  supported by the Natural Sciences and Engineering Research                               
          Council of Canada (NSERC)  \\                                                            
$^{b}$ &  supported by the FCAR of Qu\'ebec, Canada  \\                                            
$^{c}$ &  supported by the German Federal Ministry for Education and                               
          Science, Research and Technology (BMBF), under contract                                  
          numbers 057BN19P, 057FR19P, 057HH19P, 057HH29P, 057SI75I \\                              
$^{d}$ &  supported by the MINERVA Gesellschaft f\"ur Forschung GmbH, the                          
          Israel Science Foundation, the U.S.-Israel Binational Science                            
          Foundation, the Israel Ministry of Science and the Benozyio Center                       
          for High Energy Physics\\                                                                
$^{e}$ &  supported by the German-Israeli Foundation, the Israel Science                           
          Foundation, and by the Israel Ministry of Science \\                                     
$^{f}$ &  supported by the Italian National Institute for Nuclear Physics                          
          (INFN) \\                                                                                
$^{g}$ &  supported by the Japanese Ministry of Education, Science and                             
          Culture (the Monbusho) and its grants for Scientific Research \\                         
$^{h}$ &  supported by the Korean Ministry of Education and Korea Science                          
          and Engineering Foundation  \\                                                           
$^{i}$ &  supported by the Netherlands Foundation for Research on                                  
          Matter (FOM) \\                                                                          
$^{j}$ &  supported by the Polish State Committee for Scientific Research,                         
          grant no. 2P03B04616, 620/E-77/SPUB-M/DESY/P-03/DZ 247/2000 and                          
          112/E-356/SPUB-M/DESY/P-03/DZ 3001/2000\\                                                
$^{l}$ &  partially supported by the German Federal Ministry for                                   
          Education and Science, Research and Technology (BMBF)  \\                                
$^{m}$ &  supported by the Fund for Fundamental Research of Russian Ministry                       
          for Science and Edu\-cation and by the German Federal Ministry for                       
          Education and Science, Research and Technology (BMBF) \\                                 
$^{n}$ &  supported by the Spanish Ministry of Education                                           
          and Science through funds provided by CICYT \\                                           
$^{o}$ &  supported by the Particle Physics and                                                    
          Astronomy Research Council, UK \\                                                        
$^{p}$ &  supported by the US Department of Energy \\                                              
$^{q}$ &  supported by the US National Science Foundation                                          
\end{tabular}                                                                                      

                                                           %

\pagenumbering{arabic} 
\pagestyle{plain}

\section{\bf Introduction}
\label{s:Intro}

The measurements of the differential cross-section and
$F_2$ structure function in neutral current (NC) 
deep inelastic scattering (DIS) at HERA have been vital
in testing the predictions of perturbative QCD 
and in the determination
of the parton densities in the proton at low $x$.  In both of these
areas of study the accuracy and precision of the measurements are of
paramount importance.

The measurements of the double-differential
cross-section $\diff\sigma^{ep}/\diff x\diff Q^2$ and \Fem~
presented here are
based on an $e^+p$ DIS event sample twelve times larger than that used
in the previous ZEUS publication~\cite{zfp:c72:399}.
This increase, combined with an improved understanding of the
detector, results in a significant decrease in both the statistical and 
systematic uncertainties\footnote{%
Data on \Fem\ based on similar statistics and in a similar
kinematic range have recently been presented by the 
H1 collaboration~\cite{epj:c13:609,desy-00-181}.
}.  
The uncertainties are systematics dominated for
$Q^2<800$ GeV$^2$.

The outline of the paper is as follows: in Section 2, the kinematics
and the cross section of $e^+p$ NC DIS are discussed; 
in Sections 3 and 4,
the experimental conditions and the trigger used in the data taking are
presented; Section 5 explains the method of the reconstruction
of the kinematic variables; in Section 6, Monte Carlo
simulations used in the analyses are described; Sections 7 and 8
discuss the details of the basic measured quantities; in
Section 9, the characteristics of the selected events are examined;
Section 10 presents the results; and Section 11 contains the
conclusions.

\section{\bf Kinematics and cross section}
The kinematics of the deep inelastic scattering process,
$e(k) + p(P) \rightarrow e (k^\prime) + X$,
can be completely described by
the negative of the four-momentum transfer squared, $Q^2$,
and Bjorken $x$.
In the absence of QED radiation,
\begin{eqnarray}
\label{e:q2}
Q^2  & = & - q^2 =  -(k-k^\prime)^2, \nonumber \\
\label{e:x} 
x    & = & \frac{Q^2}{2P \cdot q} ,  \nonumber
\end{eqnarray}
where $k$ and $P$  are the four-momenta of the 
incoming particles and $k^\prime$ is the four-momentum of the scattered 
lepton.
The fraction of the lepton energy transferred to the proton
in its rest frame is $y = (P\cdot q)/(P\cdot k) = Q^2/(sx)$, where $s$ is the square
of the total centre-of-mass energy of the lepton-proton collision.
The invariant mass of the hadronic final state, $W$, is calculated from \begin{equation}
\label{e:w} 
W^2    =  Q^2\frac{1-x}{x} + m_p^2,    \nonumber
\end{equation}
where $m_p$ is the proton mass.

The double-differential cross section for 
inclusive $e^+p$ scattering is given in terms of the structure functions $F_i$:
\begin{eqnarray}
\label{TOTALFORM}
\frac{d^2\sigma^{e^+p}}{dx\;d\qsd} 
    &=& 
   \frac{2\pi \alpha^2}{xQ^4}
   \left[Y_+ \Ft (x,\qsd )-Y_{-}\Fd (x,\qsd )-y^2\FL (x,\qsd )\right]
   (1+\delta_r(x,\qsd ))     \\
    &=& 
\frac{d^2\sigma^{e^+p}_{\Born}}{dx\;d\qsd} 
    (1+\delta_r(x,\qsd )),      \nonumber
\end{eqnarray}
where $Y_\pm = 1 \pm (1-y)^2$ and $x$ and $Q^2$ are defined at the hadronic
vertex. In this equation, \FL\ is the longitudinal structure 
function, $xF_3$ is the parity-violating 
term arising from $Z^0$ exchange
and $\delta_r$ is the electroweak radiative correction. 
The double-differential Born cross section, $d^2\sigma_{\Born}/dxdQ^2$,
is evaluated using $1/\alpha=137.03599$, 
$\sin^2\theta_W=0.23147$ and 
$M_Z=91.1882$~GeV~\cite{epj:c15:1} 
to determine the electroweak parameters~\cite{proc:hera:1991:101}.

The structure function $F_2$ contains 
contributions from both virtual photon and $Z^0$ exchange:
\begin{equation}
\Ft  = \Fem + \frac{\qsd}{(\qsd + M_Z^2)}\Fint
+ \frac{Q^4}{(\qsd + M_Z^2)^2}\Fwk = \Fem (1+\Delta_{F_2}),
\label{eq:deltaf2}
\end{equation}
where $M_Z$ is the mass of the $Z^0$ and 
the contributions to $F_2$ from photon exchange (\Fem), 
$Z^0$ exchange (\Fwk) and the $Z^0-\gamma$ interference term (\Fint)
are separately indicated.
All three contributions are included in
the reduced cross section, $\tilde{\sigma}^{e^+p}$, defined as:
\begin{equation}
\tilde{\sigma}^{e^+p} = \Bigg[ \frac{2\pi \alpha^2}{xQ^4}Y_+\Bigg]^{-1} 
\frac{d^2\sigma^{e^+p}_{\Born}}{dx\;d\qsd}. \nonumber
\end{equation}

\section{\bf Experimental conditions}
The data presented here were taken in 1996 and 1997 using the ZEUS detector at HERA. 
In this period
HERA operated with 174-177  bunches of 820~GeV protons 
and 27.5~GeV positrons, which collided every 96~ns.
Additional unpaired positron, unpaired proton 
and empty bunches
were used to determine beam-related backgrounds. 
The proton bunch length was approximately 11 cm (r.m.s.) while 
the positron bunch length was negligible in comparison.
Variations of the mean interaction position (from run to run) lead to
an effective length of the interaction region 
of 11.5 cm (r.m.s.) centred\footnote{\ZcoosysA}
 around
$Z=-0.5$~cm.
Approximately 6\% of the proton current was contained 
in satellite bunches, which were shifted by $\pm$4.8~ns with respect 
to the nominal bunch crossing time, resulting in a fraction of 
the $ep$ interactions occurring at $\langle Z \rangle\simeq\pm 72~{\rm cm}$. 
\label{s:detector}

A brief outline of the components which are most relevant 
for this analysis, as illustrated in Fig.~\ref{fig:zeus}, 
is given below.
The uranium-scintillator calorimeter (CAL)~\citeCAL covers
99.7\% of the total solid angle. It consists of the barrel calorimeter (BCAL),
covering the polar angle range $36.7^\circ~<~\theta~<~129.1^\circ$, 
the forward calorimeter 
(FCAL), covering $2.6^\circ<\theta<36.7^\circ$ and the rear calorimeter (RCAL), covering
 $129.1^\circ<\theta<178.5^\circ$. 
The FCAL and RCAL are divided 
vertically into two halves to 
allow temporary retraction during beam injection. 
The CAL is segmented 
into electromagnetic (EMC) and hadronic (HAC) sections.
Each section is further subdivided into cells of typically 
$5 \times 20$ cm$^2$ ($10 \times 20$~cm$^2$ 
in the RCAL) for the EMC  
and $20 \times 20$~cm$^2$ for the HAC sections.
Each cell is viewed by two photomultiplier tubes (PMTs).
Under test beam conditions, the calorimeter has a single particle 
energy resolution of 
$\sigma/E$~=~18\%/$\sqrt{E ({\rm GeV})}$ 
for electrons and 
$\sigma/E$~=~35\%/$\sqrt {E({\rm GeV})}$ for hadrons~\citeCAL.
The timing resolution of a calorimeter cell is better  than 1 ns for energy
deposits greater than 4.5~\Gev .  
To minimise the effects of noise coming from the uranium radioactivity,
all EMC (HAC) cells with an energy deposit of less than 60 (110)~MeV 
were discarded in the
analysis. For the remaining cells with no neighbouring energy deposits,
this cut was increased to 100 (160)~MeV. 

A presampler (PRES)~\cite{nim:a382:419} 
is mounted in front of FCAL and RCAL. It consists
of scintillator tiles that detect 
particles originating from showers in the material between the interaction point 
and the calorimeter.
This information is used to correct the energy of the scattered positron.

The tracking system consists of 
a central tracking chamber (CTD)~\citeCTD and a rear tracking 
detector (RTD)~\cite{zeus:1993:bluebook}
enclosed in a 1.43 T solenoidal magnetic field. The interaction vertex
was measured with a typical resolution along (transverse to) the beam direction 
of 0.4~(0.1)~cm. 
The CTD was used to measure track momenta with a resolution of
$\sigma(p_t)/p_t = 0.0058p_t\oplus 0.0065\oplus0.0014/p_t$ ($p_t$ in GeV) 
and to extrapolate tracks onto the face of the
calorimeter with a resolution of $0.3\;\rm cm$.

The position of positrons scattered at small angles 
to the positron beam direction
was measured using the small angle rear tracking detector (SRTD)~\cite{nim:a401:63},
which is attached to the front face of the RCAL.
The SRTD consists of two planes of scintillator strips, 1 cm wide and 
0.5 cm thick, arranged in orthogonal directions and read out 
via optical fibres and photomultiplier tubes. 
It covers the region of about 68~$\times$~68~${\rm cm^2}$ in $X$  
and $Y$ and is positioned at $Z = -148$~cm.
A hole of 8~$\times$~20~${\rm cm^2}$ at the centre of the RCAL and SRTD
accommodates the beampipe. 
The SRTD signals resolve single minimum-ionising particles 
and provide a transverse position resolution of 3 mm.
The time resolution is better than 2 ns for a minimum-ionising particle. 
The RTD was used to determine accurately the position of the SRTD.

The hadron-electron separator (HES)~\cite{zeus:1993:bluebook} 
consists of a layer of silicon pad detectors.
The rear HES is located in the RCAL at a depth of 
3.3 radiation lengths. Each silicon pad has an area of 
$28.9\times 30.5$ mm$^2$, providing a spatial resolution of
about 9 mm for a single hit pad. If more than one adjacent pad
is hit by a shower, a cluster consisting of at most $3\times 3$ pads
around the most energetic pad is considered, 
giving a positron position resolution of 5~mm.

The luminosity was measured via the
brems\-strahlung process, $ep \rightarrow e \gamma p$,
using a lead-scintillator calorimeter (LUMI)~\citeLUMI
which detects photons at angles~$\le$~0.5~mrad with respect to the 
positron beam direction. 
 The LUMI photon calorimeter was also used to tag 
photons from initial-state radiation in DIS events.
It was positioned at $Z=-107$~m and 
had an intrinsic 
energy resolution of $\sigma/E$ = 18\%/$\sqrt {E ({\rm GeV})}$.
The carbon-lead filter, placed in front of the calorimeter
to shield it from synchrotron radiation, degraded its energy resolution
somewhat.
The position resolution was 0.2~cm in $X$ and $Y$. 
In addition, an electromagnetic calorimeter (positron tagger)
positioned at
$Z = -35$~m was used for tagging positrons scattered at very small angles.

\section{Trigger}
For the `high-$Q^2$ region' ($Q^2 \ge 30\;\rm GeV^2$), 
the data analysed correspond to an
integrated luminosity of $30.6\pm 0.5\;\rm pb^{-1}$.
For the `low-$Q^2$ region' ($Q^2 \ge 2\;\rm GeV^2$),
where the positrons are scattered through smaller angles,
the data set corresponds
to an integrated luminosity of $2.23\pm 0.02\;\rm pb^{-1}$,
which was obtained in 
short dedicated running periods.

Events were selected online by a three-level trigger 
system~\cite{zeus:1993:bluebook}.
At the first level (FLT)~\citeCFLT, 
events with a positron candidate were selected by the logical OR of the
following conditions:

\begin{itemize}
 \item {total EMC energy deposit in the BCAL greater than 4.8 GeV;}
 \item {total EMC energy deposit in the RCAL, excluding
        the ring of $20\times 20$~cm$^2$ towers 
        closest to the rear beampipe, greater than 3.4 GeV;}
 \item {an isolated positron condition (ISO-e) in the RCAL.
        The ISO-e condition requires that the isolated EMC energy deposit
        be greater than 2.08 GeV and that the HAC energy behind it
        be either less than 0.95 GeV or no more than one third of the EMC
        energy. 
        The isolation criterion required all towers adjacent to the 
        EMC energy deposit associated with the positron to have energy 
        deposits less than 2.08 GeV in the EMC and 0.95 GeV in the HAC.
        Additionally, the total
        energy deposit in CAL was required to be greater than 
        0.464~GeV and a signal above threshold
          was required in the SRTD;}
 \item  {total transverse energy in the CAL greater than 30 GeV;}
 \item  {total transverse energy in the CAL greater than 11.6 GeV
        and at least one track candidate in the CTD;}
 \item  {total EMC energy deposit in the CAL greater than 10 GeV
        and at least one track candidate in the CTD.}
\end{itemize}

Backgrounds from protons interacting outside the detector were rejected
using the time measurement of the energy deposits in upstream veto counters
and in the SRTD.  

The trigger efficiency of the FLT for events that passed
the offline selection cuts (see Section~\ref{ANA}) 
was greater than 99\%, as determined from
dedicated trigger studies 
and from Monte Carlo (MC) simulations. 

The second-level trigger (SLT)
further reduced  the background
using the times measured in the calorimeter cells and the summed energies
from the calorimeter. 
Energy and momentum conservation require
$ep$-collisions to conserve  $\delta$:
\begin{equation}
\label{eq:delta}
\delta = E-P_Z = \sum_i E_i(1-\cos\theta_i). 
\end{equation}
In this equation,
$E_i$ and $\theta_i$ are the energies and polar angles
of all energy deposits in the 
CAL.
For perfect detector resolution and for fully contained DIS events,
$\delta$ will equal twice the positron beam energy (55~\Gev ).
For photoproduction events,
where the scattered positron escapes down the beampipe,
$\delta$ peaks at much lower values. Proton beam-gas events 
that originate from inside the detector have
energy concentrated
in the forward direction and so also have small values of $\delta$.

The events were accepted at the SLT if they satisfied the condition that
\begin{equation}
  \delta_{SLT} = \sum_i E_i(1-\cos\vartheta_i) > (29~\Gev - 2E_{\gamma})\nonumber
\end{equation}
where $\vartheta_i$ is the polar angle of the energy deposit 
with respect to the nominal vertex
and $E_{\gamma}$
is the energy deposit measured in the LUMI photon calorimeter.

The full event information was available at the third-level trigger (TLT).
Tighter timing cuts as well as algorithms to remove beam-halo muons 
and cosmic rays were applied.
The quantity $\delta_{TLT}$ was determined 
using the reconstructed vertex
and the events were required to have
\mbox{$\delta_{TLT}>(30~\Gev- 2E_{\gamma})$}.
Finally, events were accepted if 
a scattered positron candidate with energy greater than 4~GeV was found.
To ensure full containment of the shower in the calorimeter, 
events in which the scattered
positron was located within 2 cm of the beamhole were rejected. 

In total 1,747,944 (3,935,516) NC DIS candidates satisfied the above 
low-$Q^2$ (high-$Q^2$) trigger condition.

\section{\bf Reconstruction of event kinematics}
\label{KINREC}

In the determination of the event kinematics, the CAL energy deposits
were separated into two categories:
those associated with the identified scattered positron, and
all other energy deposits.
The latter were defined as the hadronic energy.
The kinematics of the event was then determined from: 
\begin{itemize}
\item the energy ($E_e^{\prime}$) and polar angle ($\theta_e$) of the 
      scattered positron;
\item the hadronic energy expressed in terms of the longitudinal:
      \begin{equation} 
      \delta_h = \sum_h (E_h - \pzh)                 \nonumber
      \end{equation} 
      and transverse:
      \begin{equation} 
      \pth = \sqrt{(\sum_h \pxh)^2 +(\sum_h \pyh)^2} \nonumber
      \end{equation} 
      components, where the sums run over all energy deposits not associated with the 
      scattered positron.
\end{itemize}
The hadronic energy flow is characterised
by an angle \gammah\  defined by:
\begin{eqnarray}
\cos \gammah &=& \frac{\ppth - \delta_h^2}{\ppth + \delta_h^2} .
  \label{COSG}
\end{eqnarray}
In the quark-parton model,
the angle \gammah\ corresponds to the polar angle of the struck quark.

The following refinements in the calculation of 
$\gamma_h$ were applied in order 
to optimise the accuracy and precision of the reconstruction:
\begin{itemize}
\item 
   the transverse momentum of the hadronic system, $\pth$, 
   was replaced by the more accurately measured transverse momentum of the
   scattered positron, $\pte$; 
\item
  at low values of $\delta_h$, 
  part of the hadronic system escapes detection 
  through the forward beamhole.
  A parameterisation, determined from MC simulation, as a function of 
  the measured ratio $\pth/\pte$, $\gamma_h$ and
  $Q^2$ was used to correct $\delta_h$:
  \begin{equation}
    \delta_{h,\textrm{cor}}=\delta_h\, {\cal C}(\pth/\pte,\gamma_h,Q^2);
    \label{eq:ptcor}
  \end{equation}
\item
   in the determination of $y$ at large values of $\delta_h$,
   the information from the positron
   provides the most accurate measurement,
   $y_{e}=1-(E^\prime_e/2E_e)(1-\cos\theta_e)$.
   However, at low values of $\delta_h$, the hadronic information provides
   a more accurate measurement, 
   $y_{\textrm{cor}}=\delta_{h,\textrm{cor}}/(2E_e)$.
   The $\Sigma$-correction~\cite{nim:a361:197}
   combines $y_{\textrm{cor}}$ and $y_{e}$ such that
   \begin{equation}
      y_{\Sigma,\textrm{cor}}=
        \frac{ \delta_{h,\textrm{cor}} y_{e} 
               + \delta_e y_{\textrm{cor}}}
             {\delta_e+\delta_{h,\textrm{cor}}},         \nonumber
   \end{equation}
   where $\delta_e=E_e^{\prime}(1-\cos\theta_e)$;
\item
   finally, $\gpt$ was calculated from:
   \begin{equation}
   \cos \gpt=\frac{\ppte-\delta_{\Sigma,\textrm{cor}}^2}
   {\ppte+\delta_{\Sigma,\textrm{cor}}^2},
   \label{eq:gammapt}
   \end{equation}
   where $\delta_{\Sigma,\textrm{cor}}=2E_e\,y_{\Sigma,\textrm{cor}}$.
\end{itemize}

For the present analysis, the PT method~\cite{zfp:c72:399} was used to reconstruct
the kinematic variables of the events: 
\begin{eqnarray}
\qpt  & = &  4E_e^2 \frac{\sin \gpt(1+\cos \theta_e)}{\sin \gpt +
                          \sin \theta_e - \sin(\gpt + \theta_e)};   \nonumber \\
\xpt & = & \frac{E_e}{E_p} 
           \frac{\sin \gpt + \sin \theta_e + \sin(\gpt+\theta_e)}
                {\sin \gpt + \sin \theta_e - \sin(\gpt+\theta_e)};  \nonumber\\
\ypt & = & \frac{\qpt}{s\xpt};                                      \nonumber \\
W^2_{\textrm{PT}} & = & \qpt\frac{1-\xpt}{\xpt}+m_p^2.              \nonumber
\end{eqnarray}

This reconstruction method achieves good
resolution and minimises biases in $x$ and $Q^2$ over the full kinematic
range.

\section{Monte Carlo simulation}
\label{section:MC}

Monte Carlo (MC) event simulation was used to correct for detector
acceptance and resolution effects. The detector simulation is based on 
the GEANT program~\cite{tech:cern-dd-ee-84-1} and incorporates the 
best understanding
of the detector and the trigger, including test beam results.
Neutral current
DIS events were simulated with radiative effects using the HERACLES 4.5.2
\cite{cpc:69:155} program interfaced to the LEPTO 6.5 \cite{cpc:101:108} 
MC program by the DJANGO6 2.4 program~\cite{cpc:81:381}.
HERACLES includes O($\alpha$) leptonic
corrections for initial- and final-state radiation, and one-loop vertex and
propagator corrections.
In addition, 
leading logarithmic effects to order O($\alpha\alpha_s$) are included.
The hadronic final state was simulated using the
colour-dipole model CDMBGF~\citeCDM including all leading-order QCD 
diagrams as 
implemented in ARIADNE 4.08~\cite{cpc:71:15} for the QCD cascade and 
JETSET 7.4~\cite{cpc:82:74} for the hadronisation.
The ARIADNE model provides the best description of the 
characteristics of the 
DIS non-diffractive hadronic final state~\cite{epj:c6:239}. 
Diffractive events which are characterised by a large 
rapidity gap in the detector~\cite{epj:c6:43,zfp:c76:613}
were simulated within ARIADNE, which was interfaced to the 
RAPGAP 2.06/52~\cite{cpc:86:147}  program.  
The latter assumes that the 
struck quark belongs to a colourless state having only a small 
fraction of the proton momentum. 
The diffractive and non-diffractive samples were mixed 
as a function of $\xpt$ and $\qpt$
as determined from the $\eta_{max}$ distribution in the data,
where $\eta_{max}$
is the pseudorapidity of the most forward 
energy deposit with more than 400 MeV.
The overall normalisation was kept fixed. 
The CTEQ4D~\cite{pr:d51:4763} parton density parameterisations were used 
in the simulations.

The vertex distribution used in the simulation
was taken from DIS events having both scattered
positrons and hadrons well reconstructed by the CTD. 
The vertex finding efficiency is well described by the MC simulation,
as is shown below (see Section~\ref{sec:posangle}).

The effects of the uranium radioactivity were studied in 
randomly triggered events. The energy deposits  in these events originate 
entirely
from the calorimeter noise. 
The rate and energy distribution were simulated in the MC program.

Twelve different MC samples with varying $Q^2$ ranges were generated
with integrated luminosities increasing with $Q^2$, from
1.2 pb$^{-1}$ for $\qsd~>~0.5~\Gevsq$  to 
1500~fb$^{-1}$ for $\qsd~>~40000~\Gevsq$.
The fluctuations in the final result
due to the MC sample statistics are small.

The main source of background in the data 
comes from photoproduction interactions that led
 to the detection of a
fake scattered positron. Resolved and direct 
photoproduction events~\cite{epj:c1:109} 
were simulated using PYTHIA 5.724~\cite{cpc:82:74}, with 
the total cross
section given by the ALLM parameterisation~\cite{pl:b269:465}. 
Resolved and direct photoproduction events,
corresponding to integrated luminosities of 
0.7 and 2.5  ${\rm pb^{-1}}$, respectively, were generated with $y>0.36$.
Events with smaller $y$ values do not contribute to the photoproduction
background due to the requirement on $\delta$.

\section{Positron identification and measurement}\label{s:PIE}
The positron-identification algorithm 
is based on a neural network
using information from the CAL~\cite{nim:a365:508}.
The network separates energy deposits in the calorimeter that are due 
to electromagnetic showers from those that are of hadronic
origin by assigning a `positron-probability' value. 
If more than one positron is identified, the one with the
highest probability is taken to be the scattered positron.
Monte Carlo studies show that
the efficiency for finding the scattered positron
increases from 70\% at 8~\Gev\ to
greater than 99\% for energies above 25~\Gev . 
This was checked by using QED-Compton events.
The results obtained from the neural network were compared to 
an alternative positron-finding algorithm~\cite{epj:c11:427}.
Systematic differences
of the order of 2\% were observed at 8~\Gev, decreasing to less 
than 1\% for energies higher than 15~\Gev.

\subsection{Positron angle measurement}
\label{sec:posangle}
The scattering angle of the positron was determined 
using either its impact position on the CAL inner face and the event vertex
or from a reconstructed track in the CTD.
For $\theta_e\lsim 150^\circ$, the scattered positron registers in
the three innermost 
superlayers of the CTD and the 
parameters of the reconstructed track, matched to 
the positron impact position, were used.

For $\theta_e\gsim 150^\circ$,
the impact position of the scattered positron on the 
inner face of the calorimeter was  
reconstructed using  
other detector elements.
The SRTD information was used, when available;
otherwise the HES information was used. If neither HES 
nor SRTD information was available,
then the CAL reconstruction was used: 
\begin{itemize}
\item SRTD -- the positron position in the SRTD was 
      determined from the centres of
      gravity of the pulse-heights of the strips in the 
      $X$ and $Y$ planes of the SRTD. 
      The measured position resolution was 3~mm.  
      For a subset
      of these positrons, the position measured by the SRTD
      was compared to the result from the track measurement. 
      The resolution of the RTD-SRTD matching was 
      3~mm and systematic deviations were less than 1~mm;
\item HES -- the position in the HES was determined from the pulse-height
      sharing in the pads. The position measurement was checked 
      by matching tracks and positron candidates. 
      The resolution of the track-HES  matching was 
      5~mm with systematic deviations smaller than 1.5~mm;
\item CAL -- the position of a shower in the CAL was determined using
      the relative pulse-height of the  
      left and right PMTs and energy deposits in neighbouring
      cells. The position resolution on the face of the CAL was about 
      10~mm.  
      The alignment of the calorimeter was checked 
      by matching tracks and positron candidates in the CAL.
      The resolution of the track-CAL matching was 
      10~mm with systematic biases 
      of less than 2~mm.
\end{itemize} 

The coordinates of the event vertex were determined from tracks reconstructed 
with the CTD. The $Z$ coordinate (\zvtx) was determined on an event-by-event
basis. However, since the transverse sizes of the beams 
were smaller than the resolutions for the $X$ and $Y$ coordinates,
the average $X$ and $Y$ vertex positions per beam fill
were used. For events not having a tracking vertex, the $Z$ 
coordinate was set to $Z=0$, the nominal position of the interaction point
averaged over all runs.

The probability of finding a vertex depended on the 
number of particles detected in the tracking detectors and thus mainly 
on the hadronic angle, \gammah . Figure~\ref{fig:vtxeff} shows the 
fraction of events having a reconstructed vertex as a function of 
$\gpt$.
For $\gpt > 70^\circ$, the vertex efficiency 
was greater than 95\perc. In the low-$Q^2$ sample, it decreased to 
about 50\%  for events
with $\gpt \sim 20^\circ$. The MC simulation reproduces the observed
behaviour well. 
The drop in vertex-finding efficiency near $\gpt = 174^\circ$
results from the combination of the characteristics of the DIS final states
and the acceptance of the CTD in the rear direction. 
At low $\gpt$, the vertex-finding efficiency is higher for high-$Q^2$ events,
since the scattered positron is detected more often by the CTD.

The $Z$ coordinate of the vertex can also be determined 
from the measurements of the arrival times of
energy deposits in the FCAL. 
The corresponding resolution is $\sigma^{\textrm{timing}}_{_Z} =9$~cm.
This determination was only used for studies of systematic uncertainties.

For polar angles $17.2^\circ\lsim \theta_e\lsim 150^\circ$ and 
positron momenta $>5\;\Gev$, the tracking
efficiency was greater than 98\perc. Thus, when the 
impact position on the CAL inner face was outside a radius of
80~cm around the beampipe, corresponding to a scattering angle of 
150$^\circ$ for events with $Z=0$, 
a track was required to match the 
positron identified in the CAL and the polar angle was taken from the 
track parameters. 
A successful match required that the distance between the extrapolated
impact point of the track on the face of the CAL and the 
position determined by the CAL
was less than 10~cm. 

The resolution of  the scattering-angle measurement was 
$\sigma^{\textrm{SRTD}}_{\theta_e} \backsimeq 2.0$~mrad for positrons 
reconstructed in the SRTD, 
$\sigma^{\textrm{HES}}_{\theta_e}  \backsimeq 3.4~$mrad for positrons 
measured in the HES,
$\sigma^{\textrm{CAL}}_{\theta_e}  \backsimeq 6.8~$mrad for positrons 
measured in the CAL and 
$\sigma^{\textrm{track}}_{\theta_e}\backsimeq 2.6~$mrad for positrons 
with a matched track.

\subsection{Positron energy determination}
\label{section:CALIB}
The scattered positron loses energy in the passive material
in front of the CAL.  In the region used for this 
analysis, this corresponds to about
1.2 radiation lengths except in areas around the rear
beampipe, $\theta \gsim 170^\circ$, and the solenoid support structure, 
$130^\circ \lsim \theta \lsim 145^\circ$, where it corresponds
to 2.0 and 2.5 radiation lengths, respectively.
In the analysis, the CAL measurement of the scattered positron energy
was corrected for these energy losses.

The correction 
was determined directly from the data using the following subsamples:

\begin{itemize}
\item{at low $y$, the scattered positron energy is kinematically constrained
      to be close to the
      positron beam energy and to be primarily a function of the scattering angle.
      For these events, called kinematic peak (KP) events, 
      the mean positron energy can be determined from the scattering angle 
      ($E_e^{\prime} \approx 2E_e/(1-\cos \theta_e)$) to within
      0.5\perc.  The KP events were selected by requiring 
      $\yjb = \delta_h/(2 E_e) <0.04$~\cite{proc:epfacility:1979:391},
      and provide an energy
      calibration at $E^{\prime}_e \approx E_e = 27.5$ GeV for $\theta \gsim 135^\circ$;}
\item{the double angle (DA) method~\cite{proc:hera:1991:23,*hoeger} 
      predicts the positron energy from
      the angular information of the scattered positron and the hadrons,
      $E^\prime_e=(2E_e\sin\gamma_h)/(\sin\gamma_h+\sin\theta_e-\sin(\gamma_h+\theta_e))$.
      This relation was used to relate the energy loss in passive material to the 
      SRTD and PRES 
      signal, in the energy range $15 \lsim E_e^{\prime} \lsim 25$ GeV,
      assuming no hard QED radiation in the initial state;}
\item{for QED Compton events 
      ($ep \rightarrow e\gamma p$) observed in the main detector, the energies of 
      the positron and the photon can be predicted
      precisely from the measurement of their scattering angles since the
      transverse momentum of the scattered proton is small.
      QED Compton events provide a calibration for 
      $5 \lsim E_e^{\prime} \lsim 20$ GeV and
      $\theta \gsim 160^\circ$;}
\item{in the range  $17.2^\circ \lsim \theta \lsim 150^\circ$, 
      the momenta of the positrons
      can be independently determined from the CTD.}
\end{itemize}

For scattered positrons in the RCAL,
the correlation between the energy lost in the passive material in front
of the calorimeter and the energy deposited in the
SRTD or presampler was used to correct the calorimeter energy measurement. 
The corrections were determined using the KP and DA 
samples. The QED Compton events were used as a check on the
correction.  

The uncertainty in the energy determination, after all corrections,
was 2\% at 8~GeV, falling linearly to 1\% at 15~GeV.
In the region covered by BCAL and FCAL, $\theta \lsim 130^\circ$, 
the positron cluster energy was corrected based on material maps implemented 
in the detector simulation package.
For scattered positrons in the BCAL~\cite{epj:c11:427} and RCAL,
the corrections for non-uniformities due to geometric effects caused 
by cell and module boundaries
were also made.
Identical studies were also performed
on the event samples generated by the MC simulations. 
The resulting corrections were applied to 
the events generated in the MC simulation.

\section{Hadronic energy determination}
\label{sec:hadene}
Energy deposits coming from the hadronic final state were used to evaluate
the quantities $\delta_h$ and $\pth$.
These energy deposits, unlike those of the positron, were not
corrected for energy loss in the passive material.
Instead, 
a combination of clusters of energy deposits in the CAL and the corresponding
tracks measured in the CTD were used in determining 
the hadronic energy~\cite{tuning:phd:2001}. 
The selected calorimeter clusters and tracks are referred to 
as energy flow objects (EFOs).
The use of track information reduces the sensitivity to energy losses 
in inactive material in front of the CAL.
However, the energies of particles for which no track information was
available (e.g., because the energy was deposited by a neutral particle)
were measured using the CAL alone.

Monte Carlo studies of the calorimeter response indicated 
that the uncorrected $\gamma_h$
calculated with Eq.~(\ref{COSG}) was biased by hadronic energy coming from interactions 
in material between the primary vertex and the calorimeter or by backsplash from the 
calorimeter (albedo),
primarily  from the FCAL.
To minimise this bias, neutral clusters with energy below 3~GeV
and with polar angles larger than a $\gamma_{\textrm{max}}$, 
$\gamma_{\textrm{max}}=\gamma_h + 50^\circ$,
were removed on an event-by-event basis. 
The choice of $50^\circ$ was derived 
from the MC simulation by minimising the dependence of $\delta_h$ on 
$\gamma_{\textrm{max}}$.
For the values of $\gamma_{\textrm{max}}$ used, the biases in 
$\delta_h$ and $\pth$ were small.
The agreement of the distributions of removed energies for
different $\gamma_h$ ranges between data and MC simulation was also good.
After a first pass of cluster removal, 
the value of $\gamma_h$ was re-calculated and 
the procedure repeated until it converged, typically after two or three
iterations. 
Removing calorimeter clusters in this manner substantially improved the resolution
and reduced the bias of the quantities $\delta_h$ and $\pth$ at low values of $\delta_h$ 
(corresponding to small values of $y$)  and
left them largely unchanged for large values of $\delta_h$.

The transverse momentum of the positron, $\pte$, 
calculated using the positron energy,
corrected as described in the previous section, was compared to the
$\pth$ of the hadrons in both the MC simulation and the data.
Uncertainties in the determination of the
hadronic energy were estimated from this comparison.
The Monte Carlo simulation of
the mean $\pth/\pte$ as 
a function of \gammah\ agrees within 2\%  with the data
for the entire kinematic range covered in this paper.
Based on these comparisons, an uncertainty of $\pm$2\% was assigned
to the hadronic-energy measurement, neglecting the uncertainty
due to the angular information in the calculation
of $\pth$, which is small.

\section{Data sample}

\subsection{Event selection}
\label{ANA}
The following cuts, using the 
energy and angle of the scattered  positron,
were used both to select NC DIS events and to reject background:
\begin{itemize}
\item a positron candidate, defined as described in Section~\ref{s:PIE};
\item $E^\prime_e > 8\;\Gev $, where $E^\prime_e$ is the corrected positron energy.
      This cut ensured high and well understood 
      positron finding efficiency and suppressed background from photoproduction;
\item events with an impact point of the scattered positron on the RCAL inside
      a box of 26$\times$~14~cm$^2$ centred around the beampipe were 
      rejected (``box cut'').
      This ensured that 
      the impact point was at least 2.5~cm away from the edge of the RCAL
      and therefore guaranteed 
      full containment of the electromagnetic shower in the calorimeter.
      Events in which the scattered positron traversed
      the rear cooling pipes around the beampipe were also rejected.
      In the high-$Q^2$ sample, events with 
      an impact point of the scattered positron on the RCAL
      inside a radius of 28~cm around the beampipe were also rejected;
\item a track match for $17.2^o < \theta_e \lsim 150^\circ$.
      This condition suppressed events 
      from  cosmic rays, halo-muons, photoproduction and
      DIS events in which an electromagnetic shower was falsely identified
      as the scattered positron;
\item to reduce photoproduction background further, 
      isolated positrons were selected by 
      requiring no more than a total of 5 GeV from all  
      calorimeter cells not associated with the
      scattered positron in an $\eta-\phi$ cone of radius 0.8 centred on the
      positron.
\end{itemize}

The hadronic information was used in the following selection criteria:
\begin{itemize}
\item $38 < \delta < 65\;\Gev$, where $\delta = \sum_i (E_i-P_{Zi})$.
      Here the sum runs over both the energy flow objects 
      from the hadronic system and
      the energy deposits belonging to the
      identified positron. This cut removed events with large initial-state 
      radiation and further reduced the background from photoproduction.
      For events with a scattered positron beyond the tracking 
      acceptance ($\theta_e < 17.2^o$),
      the minimum requirement was raised to $\delta>44$ GeV;
\item $\pth/\pte > 0.3$.
      For the PT method to yield a reliable measurement of $x$ and $Q^2$,
      the loss of hadronic transverse momentum
      must be small. In some events, a substantial
      part of the current jet remained in the forward beampipe. These events, 
      produced at small \gammah, could be falsely reconstructed at large 
      \gammah\ due to the CAL noise or backsplash. This cut suppressed such events.
\end{itemize}

Finally, the following cuts were applied:
\begin{itemize}
\item $\ye < 0.95$. This condition removes events where fake positrons  are found
      in the FCAL;
\item $-50\;{\rm cm} < \zvtx < 50 \;{\rm cm}$. 
      This cut was applied
      to events with a reconstructed tracking vertex; it 
      suppressed beam-gas background
      events and the small fraction of the events in which the vertex position
      was incorrectly measured. Events without a tracking vertex 
      were accepted and were assigned 
      the mean vertex $Z$ position;
\item $\wpt>20$ GeV. This cut, corresponding
      to $y\gsim 0.004$ for $Q^2\lsim 400$ GeV$^2$,
      selected the kinematic region where 
      a reliable measurement could be made. 
      Events below this cut failed the
      acceptance and purity requirements described in 
      Section~\ref{sec:bins}.
      At low $W$,
      the hadronic system is partially lost in the forward beampipe
      and therefore the measurements are sensitive to the
      detailed simulation of the beampipe region. 
\end{itemize}
A total of 561,405 (1,300,887) events passed the low (high) $Q^2$ selection cuts.

\subsection{Binning and resolution}
\label{sec:bins}
The relative resolution in $x$ improves from 
approximately 35\% at $y\sim 0.005$, to $\sim$11\%
for $y>0.1$. The relative resolution in $Q^2$ is $\sim$2\% 
for $y\sim 0.1$ and deteriorates to approximately 5\% 
at $y\gtrsim 0.5$ and $\sim$10\% at $y\lesssim 0.01$. 
For $Q^2 > 400\;\rm GeV^2$, the $Q^2$ relative resolution
is approximately 2.5\%.

The (\x,\qsd)-bins used for the determination of 
the structure function are shown in Fig.~\ref{fig:accpu}. 
The numbers of events in each of the bins
are given in Table~\ref{tab:result}.
The bins were chosen commensurate with the 
resolutions. At large $Q^2$ and also at low \y , larger bin sizes 
were chosen
to obtain adequate statistics in each bin and to minimise bin-to-bin 
migrations as a result of the non-Gaussian tails. 
Furthermore,
good acceptance ($A>20\perc$) and purity ($P>30\perc$)
were required for each bin.
The acceptance was defined as the fraction of events generated in a bin 
that passed the event selection.
The purity was defined as the fraction of events reconstructed
in a bin that were generated in that bin.
The lowest values of acceptance occur at the lowest $Q^2$ values where the 
box cut (see Section~\ref{ANA})
becomes effective. The bins with lowest purity occur at the 
lowest \y\ values. 
In the majority of bins, the acceptance is greater than 80\% 
and the purity is
greater than 50\%, as shown in Fig.~\ref{fig:accpu}.

\subsection{Background estimate}
\begin{figure}[!b]
  \vfill
  \begin{center}
    \begin{picture}(450,200)(0,0)
    \put(-10,-10){\includegraphics[bb=10 270 540 540,scale=0.9]{%
        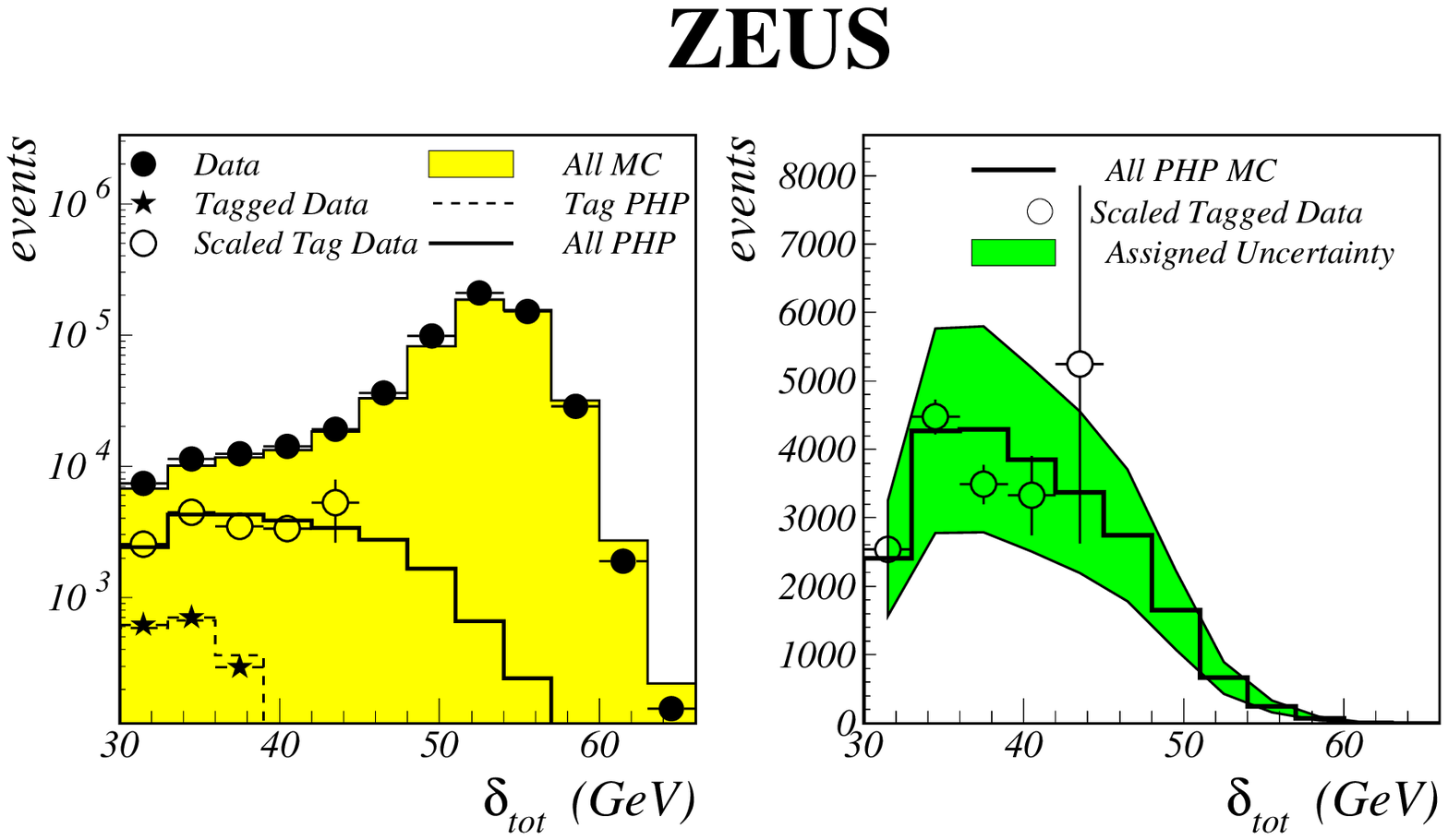}}
    \PText(45,165)(0)[c]{(a)}
    \PText(270,165)(0)[c]{(b)}
    \end{picture}
  \end{center}
      \caption[]{
  a) Distribution of $\delta$, see Eq.~(\ref{eq:delta}),
     for the low-$Q^2$ sample (filled circles),
     for that subset of the data tagged by the positron tagger (stars)
     and for these tagged data corrected for the acceptance 
     of the positron tagger (open circles).
     The lines show MC distributions after reweighting as
     discussed in the text.
  b) The photoproduction MC (PHP) and the scaled tagged data 
     are shown on a 
     linear scale.
     The shaded area shows 
     the uncertainty assigned to the photoproduction background, 
     see Section~\ref{SYSUNC}.}
  \vfill
  \label{fig:php}
  \end{figure}

The final sample contains a small number of background events
from the following sources:
\begin{itemize}
\item  {\bf non-$ep$ background.} The background not associated with
      $ep$ collisions was determined from the number of events observed in
      unpaired or empty bunches. This background was subtracted statistically,
      taking into account the appropriate ratios of bunch currents and numbers of 
      bunches. It is less than 1\% for most bins and reaches
      3\% for a few bins at low $y$;
\item  {\bf photoproduction background.} The events from the photoproduction 
      MC generator PYTHIA
      were analysed in the same way as the data and the
      number of events passing the selection cuts in each  bin 
      was determined. This number of events was subtracted.
      In the region $\delta <40\;\Gev$, the photoproduction
      contribution was measured directly through events tagged by the
      positron tagger, as seen in Fig.~\ref{fig:php}a.
      The MC photoproduction background sample was normalised to the
      measurement at $\delta<40$~GeV and subtracted bin by bin. 
      The total photoproduction background contribution to each bin is given 
      in Table~\ref{tab:result}.
      The maximum background fraction of about 20\% occurs in bins at high $y$ 
      and low $Q^2$.
\item {\bf charged current background.} The background from
      charged current interactions is negligible.
\end{itemize}

\subsection{Event characteristics}

Figure~\ref{fig:distr} shows the characteristics of both 
low- and high-$Q^2$ samples of accepted events.
To obtain these 
distributions, the MC events were reweighted to the structure function
obtained from the NLO QCD fit to these data (see Section~\ref{sec:nlofit}). 
The energy, angular and transverse momentum distributions are well reproduced 
by the MC simulation (also indicated in the figures). 

Figure~\ref{fig:kin} shows the distributions in \ypt , \xpt\ and
\qpt\, together with the MC distributions
normalised to the integrated luminosity of the data. 
Good agreement for the high-$Q^2$ sample is seen. 

The discrepancies seen at low $y$ and low $Q^2$, $Q^2\lesssim 15$~GeV$^2$
and $y\lesssim 10^{-2}$,
are located in three ($x,Q^2$)-bins, the highest $x$ bins
at $Q^2=3.5$, $6.5$ and $10$~GeV$^2$,
from a total of 242 bins.
The systematic uncertainty in this low-$y$ region (see Section~\ref{SYSUNC})
is 10\%. 
This discrepancy between data and
MC simulation is 
approximately equal
to the systematic uncertainty (not shown in the figure).
Reweighting the MC simulation to get agreement in this region
does not affect the extracted structure function values.

\section{Results}
\label{section:rew}

\subsection{Extraction of the proton structure function~$\mathbf{\Fem}$
            and the reduced cross section}
\label{sec:extract}
Monte Carlo events were generated according to Eq. (\ref{TOTALFORM})
including QED radiative effects.
The value of $\tilde{\sigma}^{e^+p}$, at fixed ($x,Q^2$) within a bin,
was obtained from the ratio
of the number of observed events to the number of events predicted from
the MC simulation in that bin,
multiplied by the reduced cross section 
from the Monte Carlo generator, $\tilde{\sigma}^{e^+p}_{\textrm{MC}}(x,Q^2)$.
The acceptance correction and bin-centering correction are thus taken from 
the MC simulation.
The MC events contain transverse and longitudinal photon and $Z^0$
contributions, which can be written as
\Fint, \Fwk, \Fd\ and \FL\, or, equivalently, the {\em relative} 
corrections $\Delta_{F_2}$ (see Eq.(\ref{eq:deltaf2})), 
$\Delta_{xF_3}$ and $\Delta_{F_L}$ according to:
\begin{equation}
\label{eq:fcor}
\tilde{\sigma}^{e^+p} = 
\Fem (1+\Delta_{F_2}+\Delta_{xF_3}+\Delta_{F_L}) = 
\Fem (1+\Delta_{all}). 
\end{equation}

The result for \Fem\ was obtained by correcting the
reduced cross section, $\tilde{\sigma}^{e^+p}$, for these
relative contributions, $\Delta_{all}$,
from the CTEQ4D parton distribution functions~\cite{pr:d55:1280}
used in the MC generation.
The reduced cross section and \Fem\ results do not significantly change
if the NLO QCD fit results, instead of CTEQ4D, are used in the 
unfolding procedure.

The statistical uncertainties on the reduced cross sections were calculated from the
number of events measured in the bins, taking into account the background
and the statistical uncertainty from the MC simulation.  
Poisson statistics was used for the two bins that contain fewer than 5 events.

The reduced cross sections and \Fem\ values were corrected for
higher-order QED radiative effects not included in HERACLES.  These 
second-order leading-logarithmic
corrections, as well as third- and higher-order
terms coming from soft-photon exponentiation,
were evaluated using the program HECTOR 1.00~\cite{cpc:94:128}. 
They lie between $-0.2\perc$ and $-0.5\perc$ and vary smoothly with $Q^2$. 
A constant correction of $-0.35\perc$ was applied.

\subsection{Systematic uncertainties}
\label{SYSUNC}

Several factors contribute to the systematic uncertainties
of \Fem\ and $\tilde{\sigma}^{e^+p}$.
The relative uncertainties in $\tilde{\sigma}^{e^+p}$
are identical to those quoted for \Fem .
Ten sources of systematic uncertainties were found to be
correlated between the different \Fem\ bins and
are denoted by \{\}
in the text and in Table~\ref{tab:error}.

The ten correlated systematic uncertainties are:
\begin{itemize}
\item[\{1\}]positron finding and efficiency.\\
The positron identification efficiency was varied within the 
uncertainty found in the QED Compton study, which is 
$\sim$2\% for positron energies $E^{\prime}_e \sim 8 $ GeV and $\sim$1\%
for energies above 15 GeV.
The effect on \Fem\ is $\sim$1\%, except in the lowest $x$ bins
where it reaches about 2\perc;

\item[\{2\}] positron scattering angle - A.\\
The uncertainty in the alignment of the different 
detector elements results in a systematic uncertainty
in the angle of the scattered positron. 
The difference between the extrapolated track
position at the face of the CAL
 and the position found from the CAL, the HES or the SRTD
shows that this uncertainty was about 2~mm. 
The largest
systematic uncertainty was obtained if the separation
between the $+X$ and $-X$ halves of the RCAL was changed by $\pm$2~mm,
resulting in an uncertainty of $\sim$ 1\% for $Q^2\lesssim 200$ GeV$^2$;

\item[\{3\}] positron scattering angle - B.\\
 For positron angles measured with the tracking detector,
the scattering angle $\theta_{e}$ was changed by the estimated 
systematic uncertainty of $\pm 1$ mrad in the track angle.
This change had a 1\% effect on the measured \Fem\
at high $Q^2$;

\item[\{4\}] positron energy scale.\\
 The systematic uncertainties in  \Fem\  due
to the  uncertainties in the absolute
calorimeter energy calibration for scattered positrons
was estimated  by changing the energy scale 
in the MC simulation: the change is 2\% at 8~GeV,  linearly decreasing to 1\% at 
15 GeV (see Section~\ref{section:CALIB}).
Systematic variations  of 1\% in \Fem\ are observed;

\item[\{5-7\}] hadronic energy measurement.\\
The hadronic energy scales of the different calorimeter 
regions in the MC simulation were changed by
$\pm$2\% in turn for the FCAL \{5\}, BCAL \{6\} and RCAL \{7\},
while leaving the positron energy scale unchanged.
The value of 2\% is based on detailed comparisons of the distributions of
the quantity $\pth/\pte$ from data and MC simulation, 
see Section~\ref{sec:hadene}.
The effect on \Fem\ was typically 2\%  in these bins at the highest or lowest $y$
and about 0.5\% elsewhere;

\item[\{8\}]hadronic energy flow - A.\\
The hadronic energy flow for diffractive events is different
from that for non-diffractive events. To investigate the sensitivity of the 
PT reconstruction method to the size of the diffractive component,
the diffractive scattering cross section in the MC simulation was reweighted
such that it 
changed by $\pm 50\%$
while leaving the
correction function ${\cal C}$ (Eq.~\ref{eq:ptcor}) unchanged.
Effects of at most 2\% at low $Q^2$ and high $y$ in \Fem\  are observed.
The variations in \Fem\ are negligible elsewhere.
This is also a check on the effect of different simulations of
the hadronic energy flow between the struck parton and the proton remnant;

\item[\{9\}] background subtractions.\\
As described above, 
the photoproduction background estimated using 
the PYTHIA MC simulation 
agreed
with the photoproduction contribution extracted from events tagged by the 
positron tagger. 
Uncertainties in the positron-tagger efficiency
and in the fake-positron
background
in the PYTHIA MC simulation led to an overall uncertainty
in the photoproduction background of $\pm 35\perc$,
as is indicated in Fig.\ref{fig:php}b.
This resulted in an uncertainty in \Fem\ which is at most 11\% for 
$y\gtrsim 0.4$ and negligible elsewhere;

\item[\{10\}]hadronic energy flow - B.\\
The hadronic angle $\gpt$ was calculated using the
energy and angle of hadronic clusters (see Section~\ref{KINREC}).
The hadronic angle can lie
inside the forward beamhole for low-$y$ events, 
since the energy deposits around the beamhole are merged into one cluster.
To estimate the uncertainty
in the cluster algorithm and the simulation of the hadronic final state
in the forward direction,
a summation over energy deposits in calorimeter cells, instead of EFOs,
was used in the hadronic reconstruction.
Systematic variations in \Fem\ up to 10\% were
observed in the lowest $W$ bins, 
leaving \Fem\ at medium and high $y$ unchanged.

\end{itemize}

The four uncorrelated systematic uncertainties are:
\begin{itemize}
\item[1)] positron finding and efficiency.\\
Varying the isolation criterion requirement of 5~GeV by
$\pm 1$~GeV around an $\eta-\phi$ cone of radius 0.8 centred on the
positron has a negligible effect on \Fem;

\item[2)] positron scattering angle.\\
Enlarging the box cut, discussed in Section~\ref{ANA}, by 1 cm
in both data and MC simulation
had an effect of $\pm$1-3\% on \Fem\ for low values of 
\qsd $\lsim 10$~GeV$^2$ and a negligible effect elsewhere;

\item[3)] hadronic energy flow - A.\\
The value of $\gamma_{\textrm{max}}$ 
(see Section~\ref{sec:hadene}) was varied by $\pm 10^\circ$,
hence varying the amount of hadronic energy removed from the event.
Only small variations in the bias of $\delta_h$ were seen in the MC simulation.
This reflects the uncertainty in the MC simulation of
low-energy deposits in the CAL and the albedo effect.
Variations up to 2\% in \Fem\ were observed in the bins at 
$y\lesssim 0.02$; elsewhere the variations were within 1\%.
In the low statistics bins at $x=0.65$ and $Q^2>1200$ GeV$^2$,
this effect increased to $\sim 10\%$;

\item[4)] hadronic energy flow - B.\\
The fraction of events removed by the $\pth/\pte$ cut is
sensitive to the amount of $\pth$ lost in the forward beampipe and thus
to the details of the jet and proton-remnant fragmentation. 
The $\pth/\pte$ cut was varied from 0.3 by $\pm 0.1$, which
resulted in changes of 1\%  at low \y .

\end{itemize}

The systematic uncertainties that did not exhibit a clear correlation
were added in quadrature and are listed in Table~\ref{tab:error}
as $\delta_{unc}$.

The satellite bunches were varied in the MC simulation by $\pm$30\% to estimate the
uncertainty in the luminosity corresponding to the events in the satellite
bunch.
Effects on the measured \Fem\ were typically 0.5-1\% and
were included in the normalisation uncertainty.

Both the uncorrelated and correlated contributions to the systematic uncertainty
were symmetrised by averaging the positive and negative
deviations. For the asymmetric systematic checks, such as the boxcut and the check on the clustering, where only a variation 
in one direction was
applied, the variation of the \Fem\ measurement was 
symmetrised by mirroring the variation.  

\begin{figure}[p]
  \vfill
  \begin{center}
    \begin{picture}(450,550)(0,0)
    \put(0,0){\mbox{\epsfig{%
          file=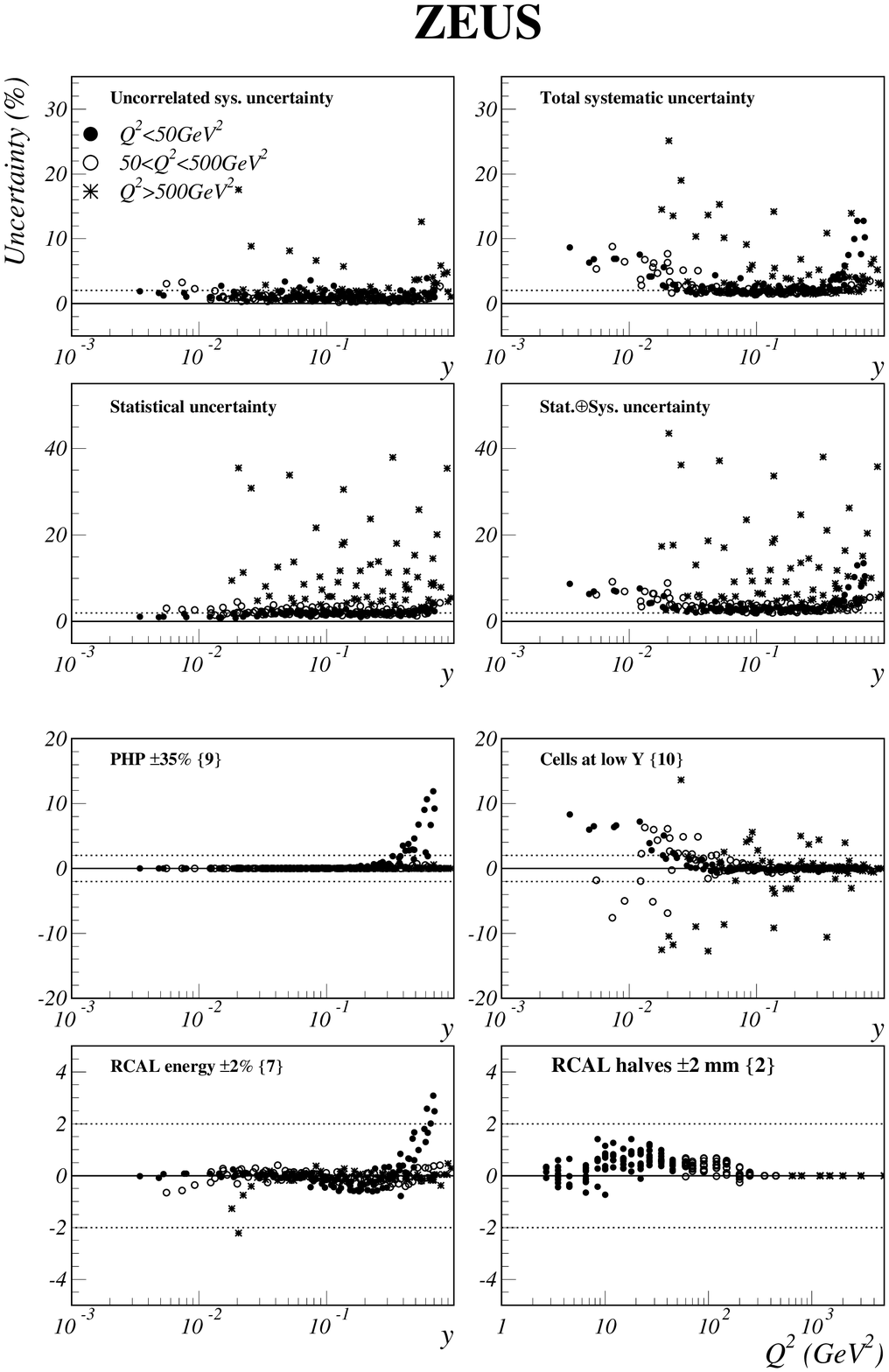,width=450pt}}}
    \PText( 60,530)(0)[c]{(a)}
    \PText(255,570)(0)[c]{(b)}
    \PText( 60,435)(0)[c]{(c)}
    \PText(255,435)(0)[c]{(d)}
    \PText( 60,278)(0)[c]{(e)}
    \PText(255,278)(0)[c]{(f)}
    \PText( 60,140)(0)[c]{(g)}
    \PText(255,140)(0)[c]{(h)}
    \end{picture}
  \end{center}
  \caption[]{
     (a-d) The relative uncorrelated systematic,
     total systematic,
     statistical, and total systematic uncertainty.
     Three regions in $Q^2$ are denoted by different symbols.
     (e-h) Four typical sources of the
     correlated systematic uncertainty, where the number 
     in \{\}
     corresponds to the number in Section~\ref{SYSUNC}.
     The dotted lines are at $\pm 2\%$.}
  \vfill
  \label{fig:errors}
  \end{figure}

Figure~\ref{fig:errors} shows the 
statistical and systematic uncertainties on \Fem\ 
for each bin as a function of \y .
Four contributions to the correlated systematic 
uncertainty are also shown.
The two largest contributions to the correlated systematic uncertainty
come from \{9\} and \{10\}.
Most sources of the systematic uncertainty show a correlation between
the points as a function of $y$. The variation of the gap
between the two RCAL halves shows a correlation in $Q^2$.
Figure~\ref{fig:error2d} shows the statistical and total
systematic uncertainty in each individual bin in the ($x,Q^2$)-plane.

The systematic uncertainties presented above do not include the uncertainty  
in the measurement of the integrated luminosity ($\pm 1.5\perc$), 
the overall trigger efficiency ($\pm 1.0\perc$), the 
higher-order electroweak radiative corrections ($\pm 0.5\perc$)
or the determination of the vertex distribution ($\pm 1.0\perc$).
These effects lead to a combined 
uncertainty of $\pm 2\perc$  on the overall normalisation of \Fem.
An additional $\pm 1\%$ relative normalisation
uncertainty was assigned to the low-$Q^2$ sample to account for the
additional uncertainty on the shape of the vertex distribution in the MC
simulation.

\subsection{Reduced cross section and \Fem }
\label{sec:pres_data}
The values of the reduced cross section $\tilde{\sigma}^{e^+p}$ and 
the electromagnetic structure function \Fem\  
are given in Table~\ref{tab:result} together with their 
statistical ($\delta_{stat}$) and systematic ($\delta_{sys}$) uncertainties, the 
corrections due 
to the $Z^0$ contribution to $F_2$, $\Delta_{F_2}$,
to the parity-violating component of the $Z^0$ exchange, $\Delta_{xF_3}$,
and to the longitudinal structure function, $\Delta_{F_L}$, see Eq.~(\ref{eq:fcor})\footnote{%
The table with the results and the uncertainities can be obtained from:
{\tt http://www-zeus.desy.de/publications.php3}
}.
The contribution of $F_L$ to $\tilde{\sigma}^{e^+p}$ 
becomes large towards large values of $y$;
the contribution of $Z^0$ exchange is substantial at large $Q^2$.

The \Fem\  values are displayed versus $x$ for fixed values of $Q^2$ in 
Figs.~\ref{fig:f2_1}-\ref{fig:f2_4}.
The structure function is measured with much improved 
precision with respect to the previous ZEUS results~\cite{zfp:c72:399}. 
Also shown is the result of the ZEUS NLO QCD fit described below.

Figures~\ref{fig:f2vsq} and \ref{fig:f2vsq2} 
show the \Fem\  values as a function of $Q^2$ for
fixed \x . Scaling violations are observed, 
which decrease as $x$ increases.
Where the data from 
the measurements presented here 
at high $x$ 
overlap the $x$ range covered by fixed target
experiments~\cite{np:b483:3,pl:b223:485,pr:d54:3006},
the agreement is good, as shown in Fig.~\ref{fig:f2vsq2}.  
The results also agree well with
the recently published data of the 
H1 collaboration~\cite{epj:c13:609,desy-00-181}
and with the recent parameterisations from 
CTEQ~\cite{epj:c12:375} and MRST~\cite{epj:c14:133}.

The \qsd -range  
has increased substantially with respect to 
previous ZEUS measurements. The ZEUS data 
in combination with fixed target data 
now span 
almost four orders of magnitude in $Q^2$ at $x\sim 0.4$.

\subsection{NLO QCD fit}
\label{sec:nlofit}
A next-to-leading-order (NLO) QCD fit to the 
reduced cross section from this analysis
has been performed.
Apart from the present data, proton-target data from
NMC~\cite{np:b483:3}, BCDMS~\cite{pl:b223:485}, 
and E665~\cite{pr:d54:3006} were used.
Deuteron-target data were used from 
NMC~\cite{np:b483:3} and E665~\cite{pr:d54:3006}, as well as  
data on the ratio $F_2^n/F_2^p$ from NMC~\cite{np:b487:3}.
The only neutrino results included in the fit are \Fd\ data from 
CCFR~\cite{zfp:c53:51}. 

To remain in the kinematic region where pQCD should be applicable,
only data with $Q^2>2.5$ GeV$^2$
were included in the fit. The resonance region 
and the region
affected by higher-twist contributions
and target mass effects ($x\gtrsim 0.1$, $Q^2\lesssim 20$ GeV$^2$) were
avoided by requiring $W^2>20$ GeV$^2$.

The parameterisations of the parton density functions (PDFs)
were assumed to have the following functional form, 
adapted from MRST~\cite{epj:c14:133}:
\begin{equation}
xf(x,Q^2_0) = Ax^{\delta}(1-x)^{\eta}(1 + \gamma x)     \nonumber
\end{equation}
and were evaluated at $Q^2_0 =  7$ GeV$^2$. The $x^{\delta}$ term 
parameterises the steep rise at low $x$, whereas $(1-x)^{\eta}$
describes the shape of the valence quarks at high $x$.

The fit was done in the $\overline{\mathrm{MS}}$ scheme 
with a variable-flavour-number {\it ansatz} 
(RT-VFN)~\cite{pr:d57:1998,*pl:b421:303}.
A charm mass of 1.35 GeV was assumed.
The strange quark distribution, $x(s(x)+\bar{s}(x))=2x\bar{s}(x)$, was fixed 
to 0.20 of the total sea, according to
measurements of CCFR~\cite{zfp:c65:189}.
The value of $\alpha_S$
was fixed to the world average value, $\alpha_S(M_Z^2) = 0.118$.

The QCD NLO
fit reproduces the data over the full kinematic range, indicating that 
NLO DGLAP evolution can give a consistent description.
The overall $\chi^2/\textrm{ndf}$ of the fit is 0.95 for 1263 data points,
where the $\chi^2$ is calculated using the statistical and total systematic
uncertainty added in quadrature. 
The $\chi^2/\textrm{ndf}$ 
for the present data is 0.84 for the 242 ZEUS data points.

\section{Conclusions}

Measurements of the 
reduced cross section and the
proton structure function, \Fem\ , have been presented
from an analysis of inelastic positron-proton neutral current scattering
data. 
The data were taken with the ZEUS detector at HERA during 1996 and 1997.
The structure function \Fem\ has now been measured 
over a substantially larger phase space than
covered by previous ZEUS measurements
and with statistical and systematic uncertainties below 2\% in most 
of the ($\x,\qsd$) region covered. 
The data cover $Q^2$ values between 2.7 and 30000~\Gevsq\ 
and $x$ values between $6\cdot 10^{-5}$ and $0.65$.

At large $x$ values
 and $Q^2$ values up to 70~\Gevsq, where the new data reach the 
\x -range covered by fixed-target experiments, good agreement with these 
experiments is found. 
There is also good agreement with the data of the 
H1 collaboration.
Strong scaling violations are observed for
$x<0.02$. The measured \x -$Q^2$ behaviour of \Fem\ can be described 
by QCD using NLO DGLAP evolution over the full kinematic range.

\section{Acknowledgements}
The strong support and encouragement of the DESY Directorate has
been invaluable. 
The experiment was made possible by the inventiveness and the diligent
efforts of the HERA machine group.  The design, construction and 
installation of the ZEUS detector have been made possible by the
ingenuity and dedicated efforts of many people from inside DESY and
from the home institutes who are not listed as authors. Their 
contributions are acknowledged with great appreciation. 
Useful conversations with 
H.~Spiesberger and R.~Thorne are also gratefully acknowledged.

\clearpage

{
\def\bibname{\Large\bf References}
\def\refname{\Large\bf References}
\pagestyle{plain}
\ifzeusbst
  \bibliographystyle{./BiBTeX/bst/l4z_default}
\fi
\ifzdrftbst
  \bibliographystyle{./BiBTeX/bst/l4z_draft}
\fi
\ifzbstepj
  \bibliographystyle{./BiBTeX/bst/l4z_epj}
\fi
\ifzbstnp
  \bibliographystyle{./BiBTeX/bst/l4z_np}
\fi
\ifzbstpl
  \bibliographystyle{./BiBTeX/bst/l4z_pl}
\fi
{\raggedright
\bibliography{./BiBTeX/user/syn.bib,%
              ./BiBTeX/bib/l4z_articles.bib,%
              ./BiBTeX/bib/l4z_books.bib,%
              ./BiBTeX/bib/l4z_conferences.bib,%
              ./BiBTeX/bib/l4z_h1.bib,%
              ./BiBTeX/bib/l4z_misc.bib,%
              ./BiBTeX/bib/l4z_old.bib,%
              ./BiBTeX/bib/l4z_preprints.bib,%
              ./BiBTeX/bib/l4z_replaced.bib,%
              ./BiBTeX/bib/l4z_temporary.bib,%
              ./BiBTeX/bib/l4z_zeus.bib}}

\providecommand{\etal}{et al.\xspace}
\providecommand{\coll}{Collab.\xspace}
\catcode`\@=11
\def\@bibitem#1{%
\ifmc@bstsupport
  \mc@iftail{#1}%
    {;\newline\ignorespaces}%
    {\ifmc@first\else.\fi\orig@bibitem{#1}}
  \mc@firstfalse
\else
  \mc@iftail{#1}%
    {\ignorespaces}%
    {\orig@bibitem{#1}}%
\fi}%
\catcode`\@=12
\begin{mcbibliography}{10}

\bibitem{zfp:c72:399}
ZEUS \coll, M.~Derrick \etal,
\newblock Z.\ Phys.{} {\bf C~72},~399~(1996)\relax
\relax
\bibitem{epj:c13:609}
H1 \coll, C.~Adloff \etal,
\newblock Eur.\ Phys.\ J.{} {\bf C~13},~609~(2000)\relax
\relax
\bibitem{desy-00-181}
H1 \coll, C.~Adloff \etal,
\newblock Preprint \mbox{DESY-00-181} (\mbox{hep-ex/0012053}), 2000.
\newblock Accepted by Eur.~Phys.~J.~C\relax
\relax
\bibitem{epj:c15:1}
Particle Data Group, D.E. Groom \etal,
\newblock Eur.\ Phys.\ J.{} {\bf C15},~73~(2000)\relax
\relax
\bibitem{proc:hera:1991:101}
J.~Bl\"umlein and M.~Klein,
\newblock in {\em Proc.\ Workshop on Physics at HERA, Oct.~1991},
  eds.~W.~Buchm\"uller and G.~Ingelman, Vol.~1, p.~101.
\newblock Hamburg, Germany, DESY, 1992\relax
\relax
\bibitem{nim:a309:77}
M.~Derrick \etal,
\newblock Nucl.\ Instr.\ and Meth.{} {\bf A~309},~77~(1991)\relax
\relax
\bibitem{nim:a309:101}
A.~Andresen \etal,
\newblock Nucl.\ Instr.\ and Meth.{} {\bf A~309},~101~(1991)\relax
\relax
\bibitem{nim:a321:356}
A.~Caldwell \etal,
\newblock Nucl.\ Instr.\ and Meth.{} {\bf A~321},~356~(1992)\relax
\relax
\bibitem{nim:a336:23}
A.~Bernstein \etal,
\newblock Nucl.\ Instr.\ and Meth.{} {\bf A~336},~23~(1993)\relax
\relax
\bibitem{nim:a382:419}
A.~Bamberger \etal,
\newblock Nucl.\ Instr.\ and Meth.{} {\bf A~382},~419~(1996)\relax
\relax
\bibitem{nim:a279:290}
N.~Harnew \etal,
\newblock Nucl.\ Instr.\ and Meth.{} {\bf A~279},~290~(1989)\relax
\relax
\bibitem{npps:b32:181}
B.~Foster \etal,
\newblock Nucl.\ Phys.\ Proc.\ Suppl.{} {\bf B~32},~181~(1993)\relax
\relax
\bibitem{nim:a338:254}
B.~Foster \etal,
\newblock Nucl.\ Instr.\ and Meth.{} {\bf A~338},~254~(1994)\relax
\relax
\bibitem{zeus:1993:bluebook}
ZEUS \coll, U.~Holm~(ed.),
\newblock {\em The {ZEUS} Detector},
\newblock Status Report, (unpublished), DESY, 1993,
\newblock available on
  \texttt{http://www-zeus.desy.de/bluebook/bluebook.html}\relax
\relax
\bibitem{nim:a401:63}
A.~Bamberger \etal,
\newblock Nucl.\ Instr.\ and Meth.{} {\bf A~401},~63~(1997)\relax
\relax
\bibitem{desy-92-066}
J.~Andruszk\'ow \etal,
\newblock Report \mbox{DESY-92-066}, DESY, 1992\relax
\relax
\bibitem{zfp:c63:391}
ZEUS \coll, M.~Derrick \etal,
\newblock Z.\ Phys.{} {\bf C~63},~391~(1994)\relax
\relax
\bibitem{nim:a355:278}
W.H.~Smith \etal,
\newblock Nucl.\ Instr.\ and Meth.{} {\bf A~355},~278~(1995)\relax
\relax
\bibitem{nim:a361:197}
U.~Bassler and G.~Bernardi,
\newblock Nucl.\ Instr.\ and Meth.{} {\bf A~361},~197~(1995)\relax
\relax
\bibitem{tech:cern-dd-ee-84-1}
R.~Brun et al.,
\newblock {\em {\sc geant3}},
\newblock Technical Report CERN-DD/EE/84-1, CERN, 1987\relax
\relax
\bibitem{cpc:69:155}
A.~Kwiatkowski, H.~Spiesberger and H.-J.~M\"ohring,
\newblock Comp.\ Phys.\ Comm.{} {\bf 69},~155~(1992).
\newblock Also in {\it Proc.\ Workshop Physics at HERA}, 1991, DESY,
  Hamburg\relax
\relax
\bibitem{cpc:101:108}
G.~Ingelman, A.~Edin and J.~Rathsman,
\newblock Comp.\ Phys.\ Comm.{} {\bf 101},~108~(1997)\relax
\relax
\bibitem{cpc:81:381}
K.~Charchula, G.A.~Schuler and H.~Spiesberger,
\newblock Comp.\ Phys.\ Comm.{} {\bf 81},~381~(1994)\relax
\relax
\bibitem{pl:b165:147}
Y.~Azimov \etal,
\newblock Phys.\ Lett.{} {\bf B~165},~147~(1985)\relax
\relax
\bibitem{np:b306:746}
G.~Gustafson and U.~Petterson,
\newblock Nucl.\ Phys.{} {\bf B306},~746~(1988)\relax
\relax
\bibitem{zfp:c43:625}
B.~Andersson \etal,
\newblock Z.\ Phys.{} {\bf C~43},~625~(1989)\relax
\relax
\bibitem{cpc:71:15}
L.~L\"onnblad,
\newblock Comp.\ Phys.\ Comm.{} {\bf 71},~15~(1992)\relax
\relax
\bibitem{cpc:82:74}
T.~Sj\"ostrand,
\newblock Comp.\ Phys.\ Comm.{} {\bf 82},~74~(1994)\relax
\relax
\bibitem{epj:c6:239}
ZEUS \coll, J.~Breitweg \etal,
\newblock Eur.\ Phys.\ J.{} {\bf C~6},~239~(1999)\relax
\relax
\bibitem{epj:c6:43}
ZEUS \coll, J.~Breitweg \etal,
\newblock Eur.\ Phys.\ J.{} {\bf C~6},~43~(1999)\relax
\relax
\bibitem{zfp:c76:613}
H1 \coll, C.~Adloff \etal,
\newblock Z.\ Phys.{} {\bf C~76},~613~(1997)\relax
\relax
\bibitem{cpc:86:147}
H.~Jung,
\newblock Comp.\ Phys.\ Comm.{} {\bf 86},~147~(1995)\relax
\relax
\bibitem{pr:d51:4763}
H.L.~Lai \etal,
\newblock Phys.\ Rev.{} {\bf D~51},~4763~(1995)\relax
\relax
\bibitem{epj:c1:109}
ZEUS \coll, J.~Breitweg \etal,
\newblock Eur.\ Phys.\ J.{} {\bf C~1},~109~(1998)\relax
\relax
\bibitem{pl:b269:465}
H.~Abramowicz \etal,
\newblock Phys.\ Lett.{} {\bf B~269},~465~(1991)\relax
\relax
\bibitem{nim:a365:508}
H.~Abramowicz, A.~Caldwell and R.~Sinkus,
\newblock Nucl.\ Instr.\ and Meth.{} {\bf A~365},~508~(1995)\relax
\relax
\bibitem{epj:c11:427}
ZEUS \coll, J.~Breitweg \etal,
\newblock Eur.\ Phys.\ J.{} {\bf C~11},~427~(1999)\relax
\relax
\bibitem{proc:epfacility:1979:391}
F.~Jacquet and A.~Blondel,
\newblock in {\em Proceedings of the Study for an $ep$ Facility for {Europe}},
  ed.~U.~Amaldi, p.~391.
\newblock Hamburg, Germany, 1979.
\newblock Also in preprint \mbox{DESY 79/48}\relax
\relax
\bibitem{proc:hera:1991:23}
S.~Bentvelsen, J.~Engelen and P.~Kooijman,
\newblock in {\em Proc.\ Workshop on Physics at HERA, Oct.~1991},
  eds.~W.~Buchm\"uller and G.~Ingelman, Vol.~1, p.~23.
\newblock Hamburg, Germany, DESY, 1992\relax
\relax
\bibitem{hoeger}
K.C.~H\"oger,
\newblock {\em Measurement of $x$, $y$ and $Q^2$ in neutral current events}.
\newblock Ibid, p.43\relax
\relax
\bibitem{tuning:phd:2001}
N.~Tuning,
\newblock {\em Proton structure functions at {HERA}}.
\newblock Ph.D.\ Thesis, University of Amsterdam, 2001,
\newblock available on
  \texttt{http://www.nikhef.nl/pub/services/library.html}\relax
\relax
\bibitem{pr:d55:1280}
H.L.~Lai \etal,
\newblock Phys.\ Rev.{} {\bf D~55},~1280~(1997)\relax
\relax
\bibitem{cpc:94:128}
A.~Arbuzov \etal,
\newblock Comp.\ Phys.\ Comm.{} {\bf 94},~128~(1996)\relax
\relax
\bibitem{np:b483:3}
NMC \coll, M.~Arneodo \etal,
\newblock Nucl.\ Phys.{} {\bf B~483},~3~(1997)\relax
\relax
\bibitem{pl:b223:485}
BCDMS \coll, A.C.~Benvenuti \etal,
\newblock Phys.\ Lett.{} {\bf B~223},~485~(1989)\relax
\relax
\bibitem{pr:d54:3006}
E665 \coll, M.R.~Adams \etal,
\newblock Phys.\ Rev.{} {\bf D~54},~3006~(1996)\relax
\relax
\bibitem{epj:c12:375}
CTEQ \coll, H.L.~Lai \etal,
\newblock Eur.\ Phys.\ J.{} {\bf C~12},~375~(2000)\relax
\relax
\bibitem{epj:c14:133}
A.D.~Martin \etal,
\newblock Eur.\ Phys.\ J.{} {\bf C~14},~133~(2000)\relax
\relax
\bibitem{np:b487:3}
NMC \coll, M.~Arneodo \etal,
\newblock Nucl.\ Phys.{} {\bf B~487},~3~(1997)\relax
\relax
\bibitem{zfp:c53:51}
CCFR \coll, E.~Oltman \etal,
\newblock Z.\ Phys.{} {\bf C~53},~51~(1992)\relax
\relax
\bibitem{pr:d57:1998}
R.S.~Thorne and R.G.~Roberts,
\newblock Phys.\ Rev.{} {\bf D~57},~1998~(1998)\relax
\relax
\bibitem{pl:b421:303}
R.S.~Thorne and R.G.~Roberts,
\newblock Phys.\ Lett.{} {\bf B~421},~303~(1998)\relax
\relax
\bibitem{zfp:c65:189}
CCFR \coll, A.O.~Bazarko \etal,
\newblock Z.\ Phys.{} {\bf C65},~189~(1995)\relax
\relax
\end{mcbibliography}
}
\vfill\eject
                                                 
\small
\begin{center}
\tablecaption{Results for the reduced cross section and the structure function \Fem.
The relative corrections in the last three columns are defined in Section~\ref{sec:extract}.}
\label{tab:result}
\tablefirsthead{
    \hline
    $Q^2$ & $x$ & $y$ & $N_{data}$ &  $N_{php}$ &
    $\tilde{\sigma}^{e^+p}$ & $\delta_{stat}$ & $\delta_{sys}$ & 
    $F_2^{em}$ & 
    $\Delta_{all}$ & $\Delta_{F_2}$ & $\Delta_{xF_3}$ &$\Delta_{F_L}$ \\
     \footnotesize{(GeV$^2$)} & & & & & & (\%) &  (\%) &  &  (\%) &  (\%) &  (\%) &  (\%) \\
   \hline \hline}
\tablehead{
    \multicolumn{13}{l}{
    {\normalsize {\bf Table \thetable\ } (continued): {\it 
     Results for the reduced cross section and the structure function \Fem.
     }}} \\  
    \hline
    $Q^2$ & $x$ & $y$ &  $N_{data}$ &  $N_{php}$ &
    $\tilde{\sigma}^{e^+p}$ & $\delta_{stat}$ & $\delta_{sys}$ & 
    $F_2^{em}$ & 
    $\Delta_{all}$ & $\Delta_{F_2}$ & $\Delta_{xF_3}$ &$\Delta_{F_L}$ \\
     \footnotesize{(GeV$^2$)}  & & & & & & (\%) &  (\%) &  &  (\%) &  (\%) &  (\%) &  (\%) \\
   \hline \hline}
\tabletail{\hline}

\begin{supertabular}{|r|c|c||r|r||c||r|r||c|r|r|r|r|}
 \hline
     2.7&  6.32$\cdot10^{-5}$&  0.474&  9506& 769&  0.832&   1.7&   4.7&  0.869&  -4.3&   0.0&   0.0&  -4.3\\
     2.7&  1.61$\cdot10^{-4}$&  0.186& 21489& 229&  0.740&   1.0&   2.4&  0.744&  -0.6&   0.0&   0.0&  -0.6\\
     2.7&  4.00$\cdot10^{-4}$&  0.075& 18716&  47&  0.649&   1.1&   3.9&  0.650&  -0.1&   0.0&   0.0&  -0.1\\
     2.7&  6.32$\cdot10^{-4}$&  0.047&  8745&  13&  0.607&   1.5&   4.3&  0.607&   0.0&   0.0&   0.0&   0.0\\
     2.7&  2.00$\cdot10^{-3}$&  0.015& 23981&   6&  0.477&   0.9&   4.2&  0.477&   0.0&   0.0&   0.0&   0.0\\
 \hline
     3.5&  6.32$\cdot10^{-5}$&  0.614&  5960&1422&  1.076&   2.7&  12.7&  1.177&  -8.6&   0.0&   0.0&  -8.6\\
     3.5&  1.02$\cdot10^{-4}$&  0.380&  6859& 354&  0.951&   2.0&   3.6&  0.979&  -2.9&   0.0&   0.0&  -2.9\\
     3.5&  1.61$\cdot10^{-4}$&  0.241&  5818&  62&  0.882&   2.0&   2.5&  0.891&  -1.0&   0.0&   0.0&  -1.0\\
\end{supertabular}
\end{center}

                                                
\clearpage
\scriptsize
\begin{center}
\tablecaption{Relative uncertainties in the reduced cross section. 
The systematic uncertainties, denoted by the \{\}, 
are explained in Section~\ref{SYSUNC}.
The overall normalisation uncertainty of 2\% is not included,
neither is the additional uncertainty of 1\% at low $Q^2$.}
\label{tab:error}
\tablefirsthead{
    \hline
    $Q^2$ & $x$ & 
    $\tilde{\sigma}^{e^+p}$ & $\delta_{stat}$ & $\delta_{sys}$ & 
    $\delta_{unc}$   & $\delta_{\{1\}}$ & $\delta_{\{2\}}$ &$\delta_{\{3\}}$ &
    $\delta_{\{4\}}$ & $\delta_{\{5\}}$ & $\delta_{\{6\}}$ &$\delta_{\{7\}}$ &
    $\delta_{\{8\}}$ & $\delta_{\{9\}}$ & $\delta_{\{10\}}$ \\
    \tiny{(GeV$^2$)} & &  &
    (\%)&(\%)&(\%)&(\%)&(\%)&(\%)&(\%)&(\%)&(\%)&(\%)&(\%)&(\%)&  (\%) \\
    \hline
    \hline
}
\tablehead{
    \multicolumn{16}{l}{
    {\normalsize {\bf Table \thetable\ }(continued): {\it 
     Relative uncertainties in the reduced cross section. 
     }}} \\  
    \hline
    $Q^2$ & $x$ & 
    $\tilde{\sigma}^{e^+p}$ & $\delta_{stat}$ & $\delta_{sys}$ & 
    $\delta_{unc}$   & $\delta_{\{1\}}$ & $\delta_{\{2\}}$ &$\delta_{\{3\}}$ &
    $\delta_{\{4\}}$ & $\delta_{\{5\}}$ & $\delta_{\{6\}}$ &$\delta_{\{7\}}$ &
    $\delta_{\{8\}}$ & $\delta_{\{9\}}$ & $\delta_{\{10\}}$ \\
    \tiny{(GeV$^2$)} & &  &
    (\%)&(\%)&(\%)&(\%)&(\%)&(\%)&(\%)&(\%)&(\%)&(\%)&(\%)&(\%)&  (\%) \\
    \hline
    \hline
}
\tabletail{\hline}


 \end{center}
\caption{Schematic view of the ZEUS detector
         showing those components important for this analysis.}
\label{fig:zeus}
\end{figure}

  \begin{figure}[!ht]
  \vfill
  \begin{center}
    %
  \end{center}
      \caption[]{Vertex finding efficiency as a function of the hadronic angle
   $\gpt$, defined in Eq.~(\ref{eq:gammapt}).
  }
  \vfill
  \label{fig:vtxeff}
  \end{figure}

\plotup{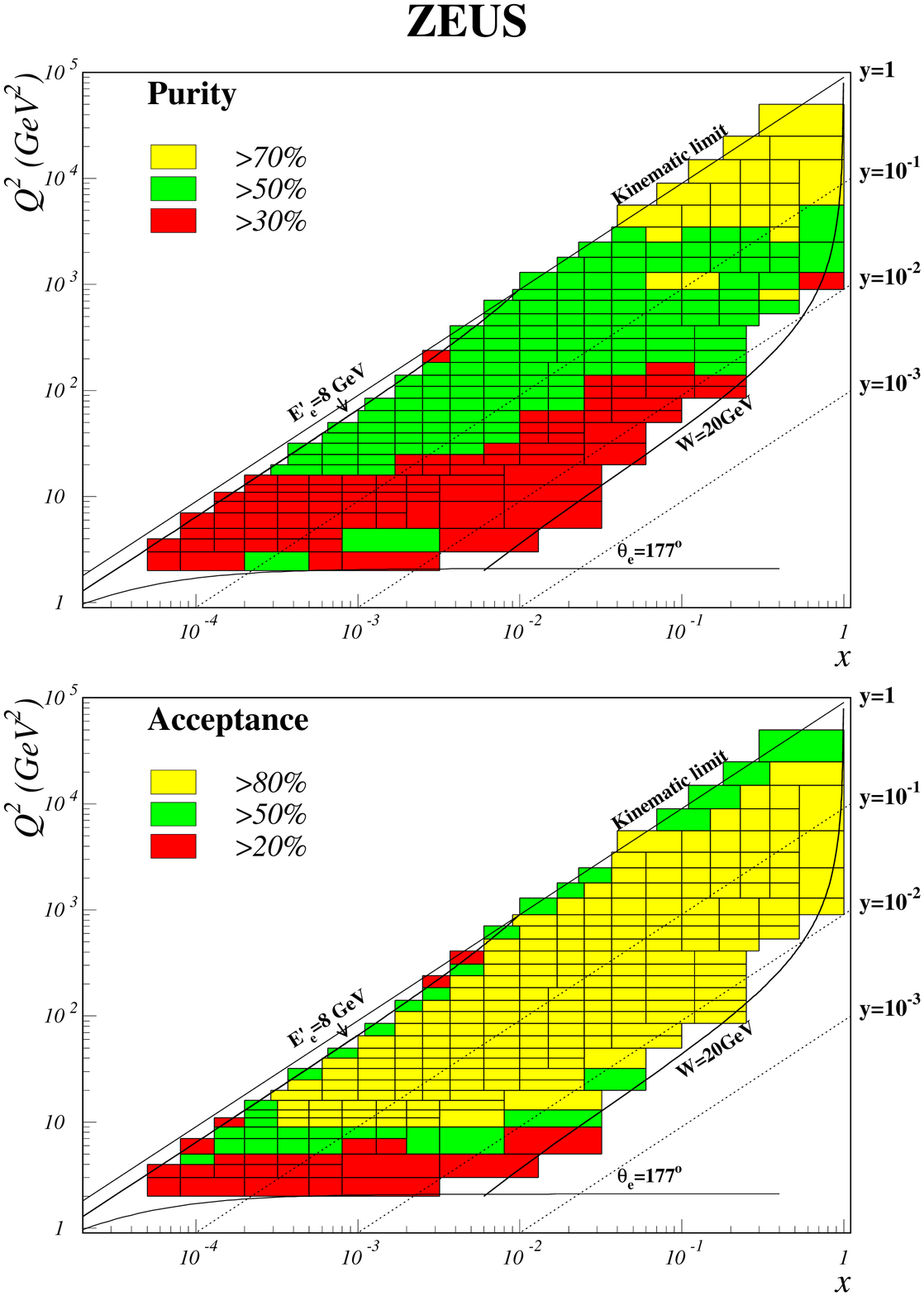}{fig:accpu}{
  a) The purity.
  b) The acceptance.
  The analysis binning is shown, 
  together with the limits of the ZEUS acceptance.
  At the edges of the
  kinematic plane, the acceptance is lower due to the various selection 
  criteria. The rear beamhole limits the acceptance of events with small 
  angles of the scattered positron to $Q^2\gtrsim 2$~GeV$^2$. 
  For $y\gtrsim 0.7$, the positron energy cut and 
  for $Q^2\gtrsim 1000$~GeV$^2$ and $y=1$ the
  kinematic limit of HERA constrains the measurements. 
  The forward beamhole 
  limits the acceptance to $y\gtrsim 0.004$ for events with small
  angles of the hadronic system; 
  this limit is shown by the line $W=20$ GeV.
}

\plot{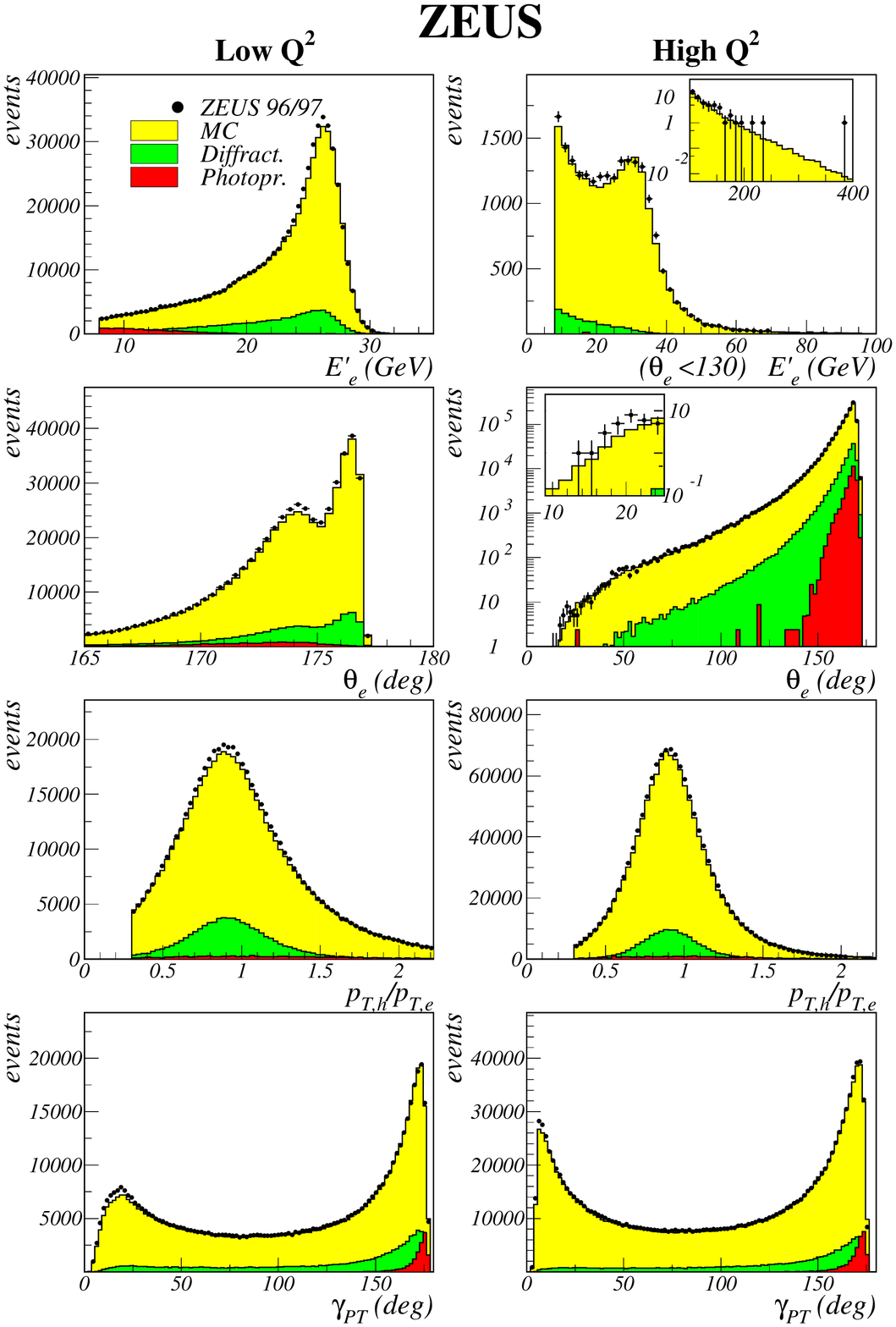}{0}{fig:distr}
{Distributions of measured quantities, as discussed in the text,
are shown for the low-$Q^2$ sample
and the high-$Q^2$ sample. 
The MC histograms 
are based on the NLO QCD fit to the data and
are normalised according to the luminosity.
The diffractive component and the photoproduction background
contributions are separately shown.
The inserts show details of the corresponding distributions.}

\plot{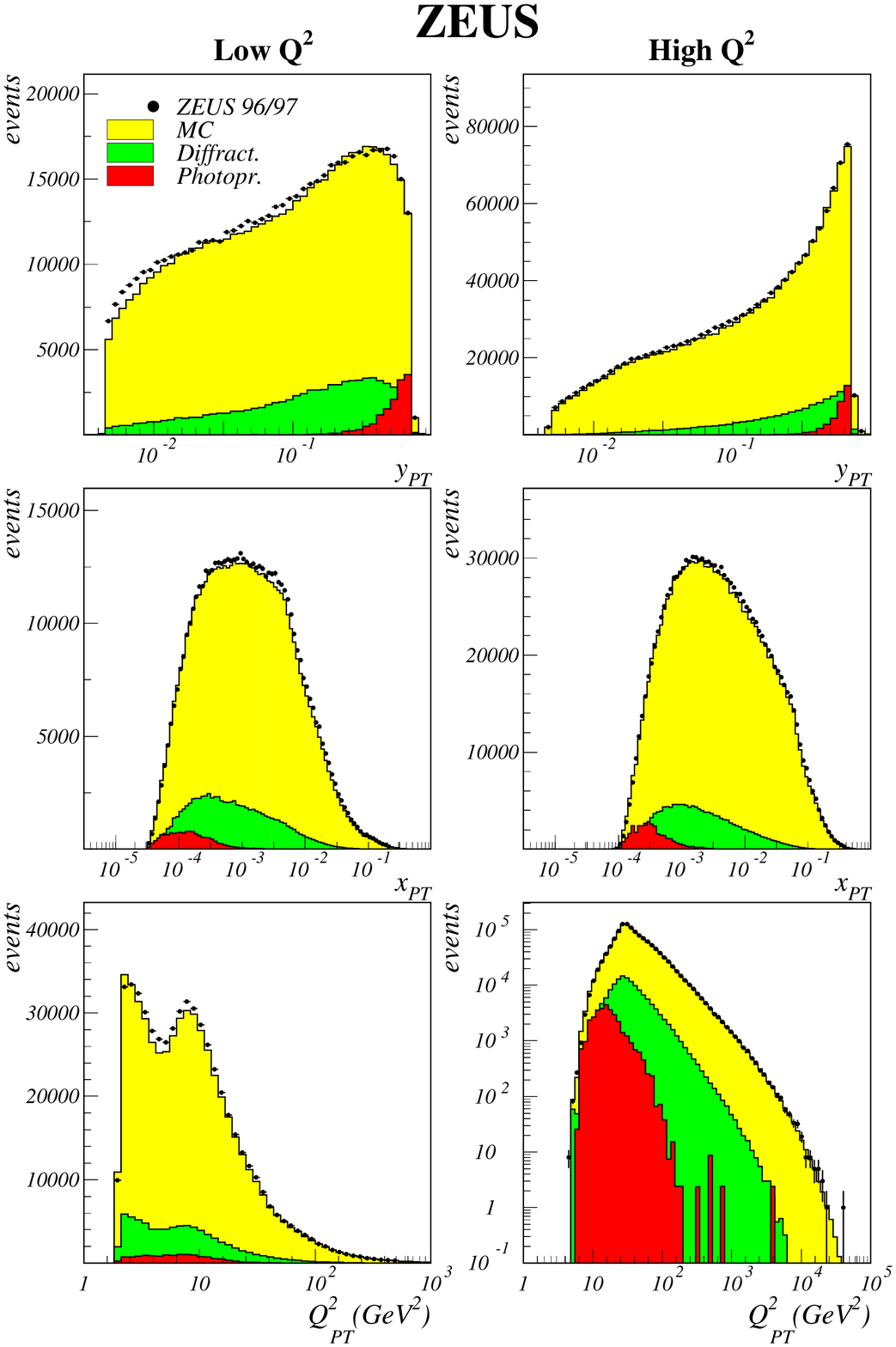}{0}{fig:kin}
{Distributions of kinematic variables
are shown for the low-$Q^2$ sample
and the high-$Q^2$ sample. 
The MC histograms 
are based on the NLO QCD fit to the data and
are normalised according to the luminosity.
The diffractive component and the photoproduction background
contributions are separately shown.
}

\plotup{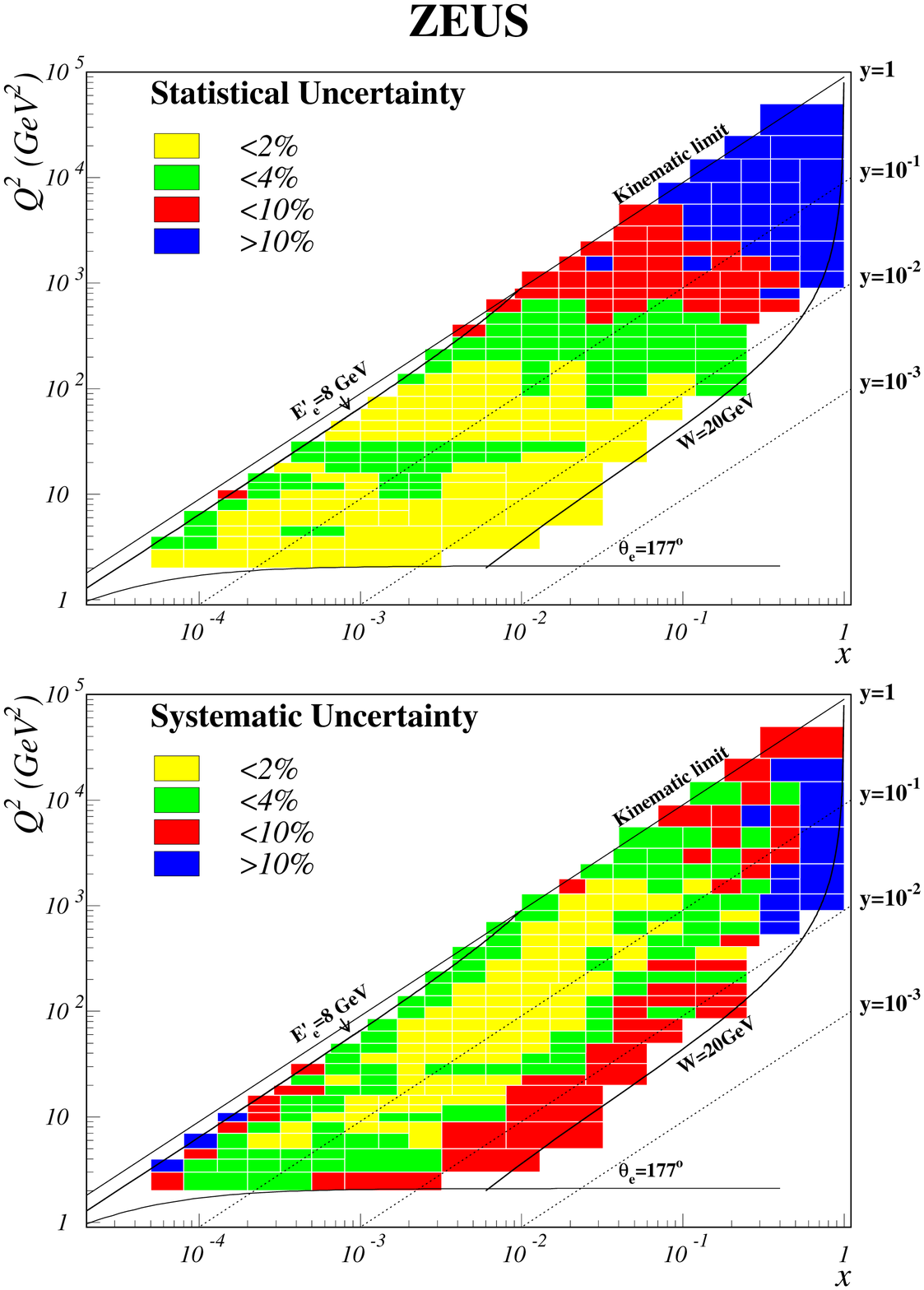}{fig:error2d}{
a) The statistical uncertainty.
b) The systematic uncertainty.
Over a large region of phase space both  the statistical and systematic
uncertainties are below 2\%. At the edges of the measurable region the
systematic uncertainty increases to $\sim 10\%$.
For $y\gtrsim 0.5$ the uncertainty increases due to the
uncertainty on the photoproduction background,
whereas for $y\lesssim 0.01$ the uncertainty increases mainly because of
uncertainties on the hadronic energy flow.
}


\plot{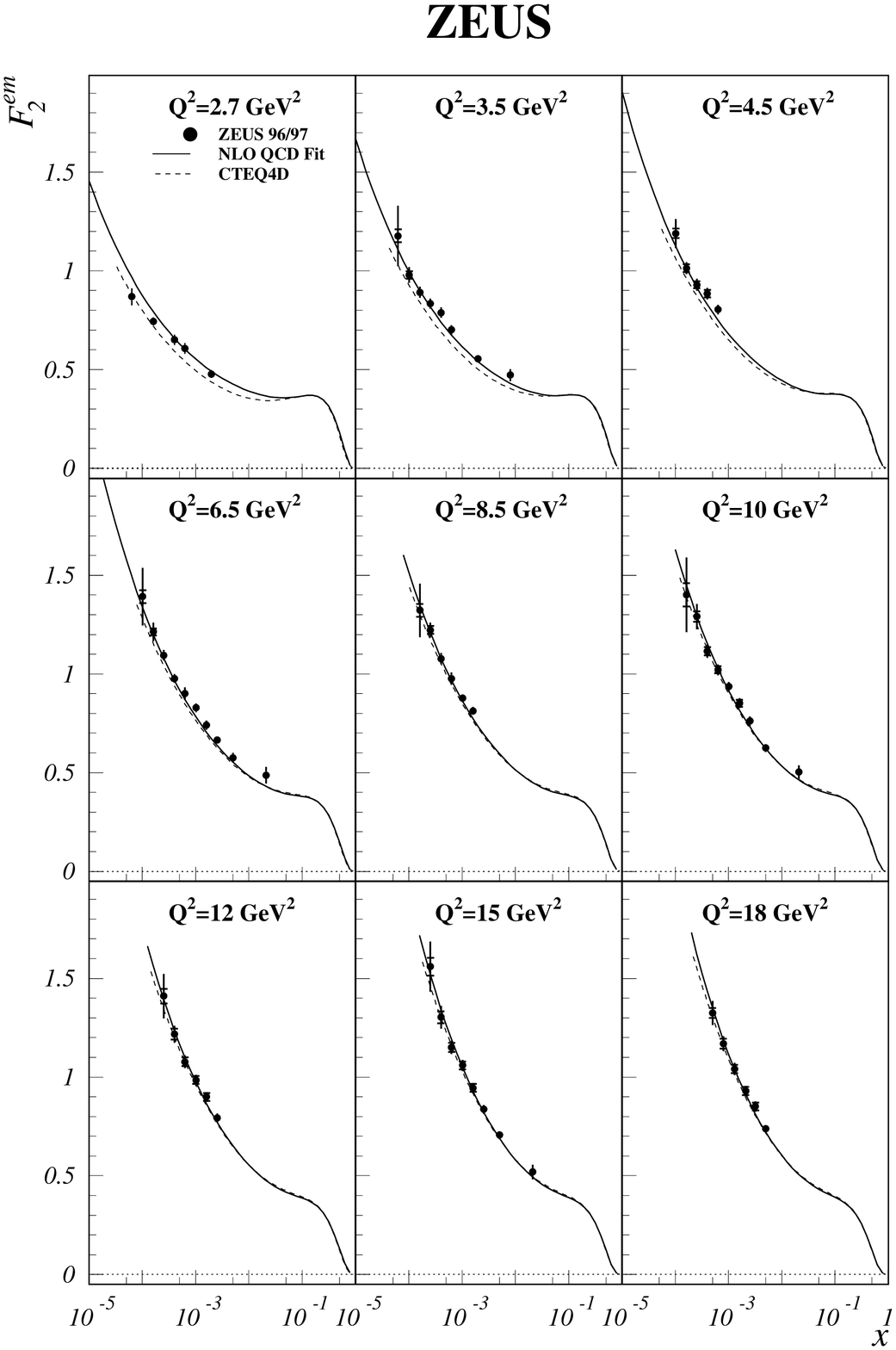}{0}{fig:f2_1}{
The results for $F_2^{em}$ (points) are shown versus $x$ for fixed $Q^2$.
The full curves show results from the ZEUS NLO QCD fit and the dashed
curves show predictions from CTEQ4D.
The inner error bars (delimited by the horizontal lines)
show the statistical uncertainties; the outer ones show the statistical and
systematic uncertainties added in quadrature.
The overall normalisation uncertainties, see Section~\ref{SYSUNC},
are not included.}

\plot{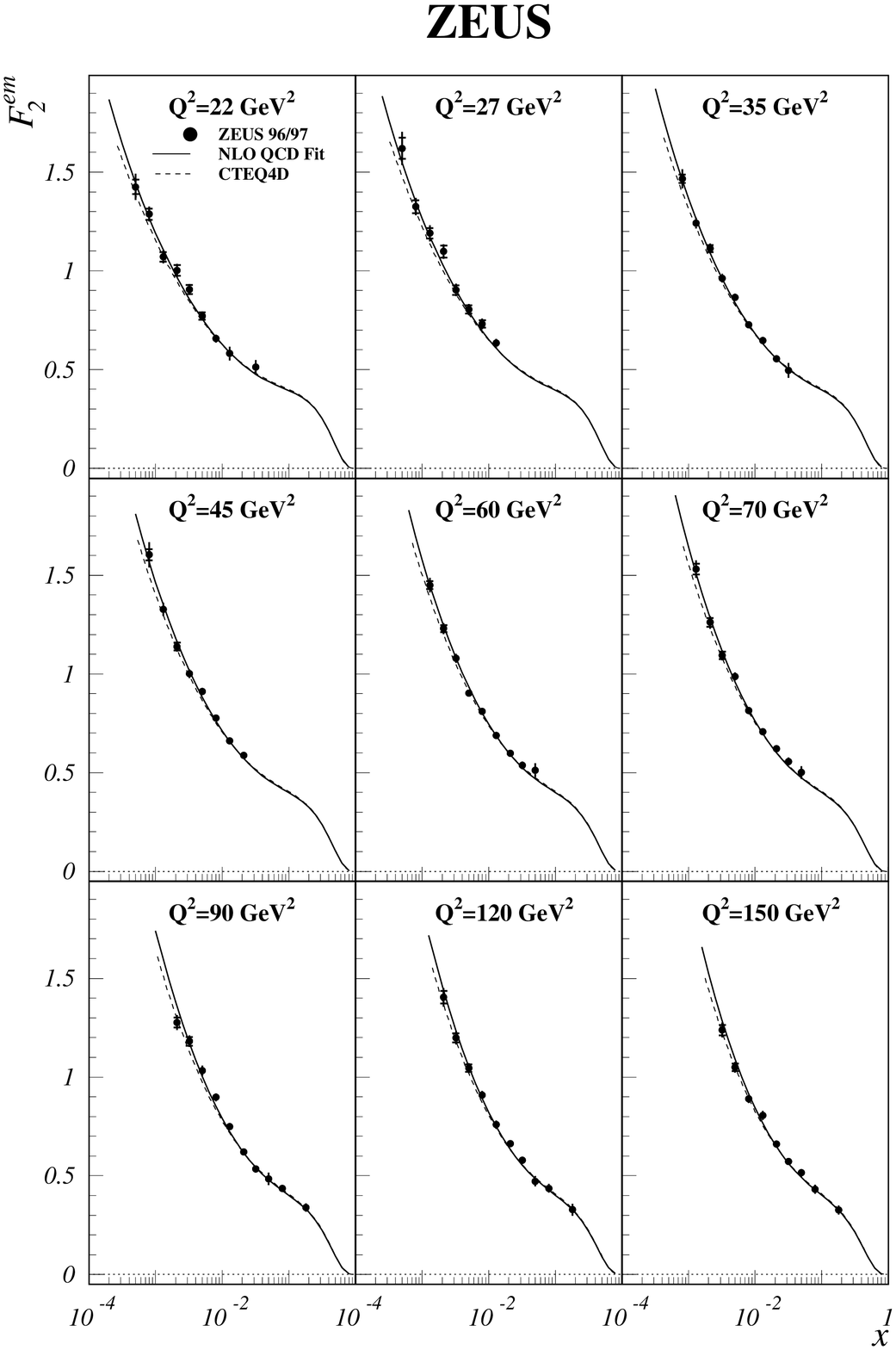}{0}{fig:f2_2}{
The results for $F_2^{em}$ (points) are shown versus $x$ for fixed $Q^2$.
For details, see caption of Fig.~\ref{fig:f2_1}.}

\plot{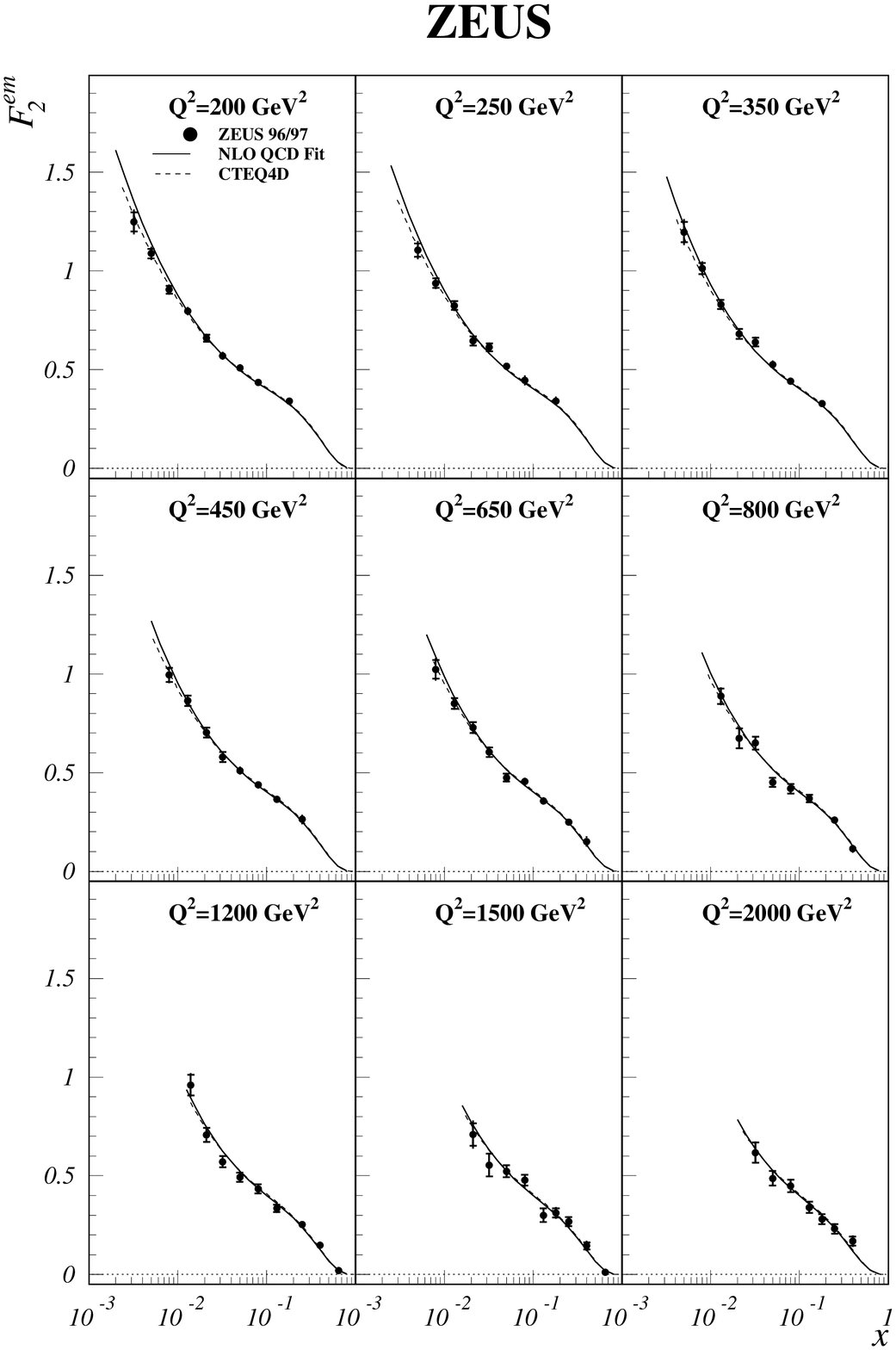}{0}{fig:f2_3}{
The results for $F_2^{em}$ (points) are shown  versus $x$ for fixed $Q^2$.
For details, see caption of Fig.~\ref{fig:f2_1}.}

\plot{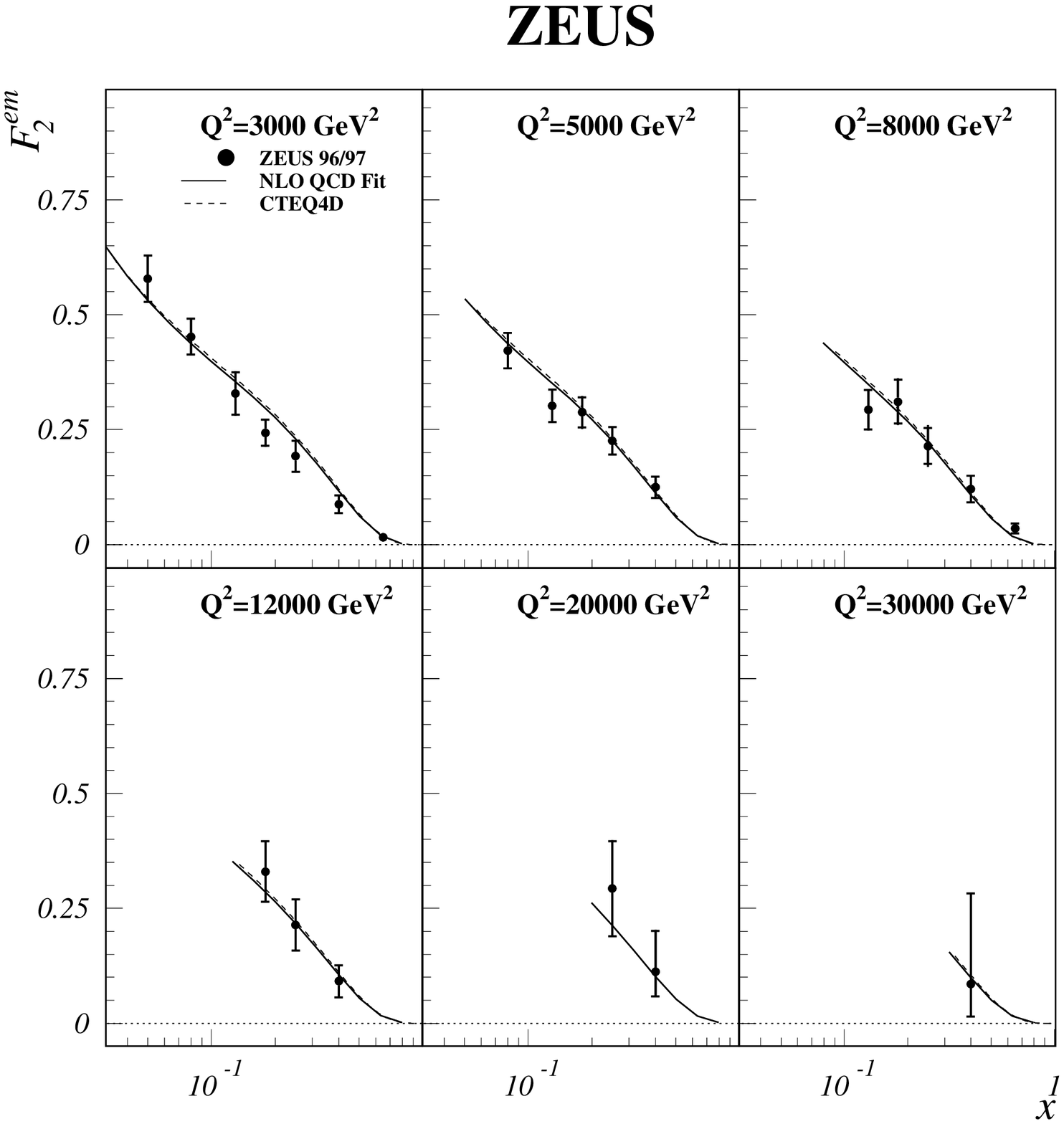}{0}{fig:f2_4}{
The results for $F_2^{em}$ (points) are shown versus $x$ for fixed $Q^2$.
For details, see caption of Fig.~\ref{fig:f2_1}.}

\plot{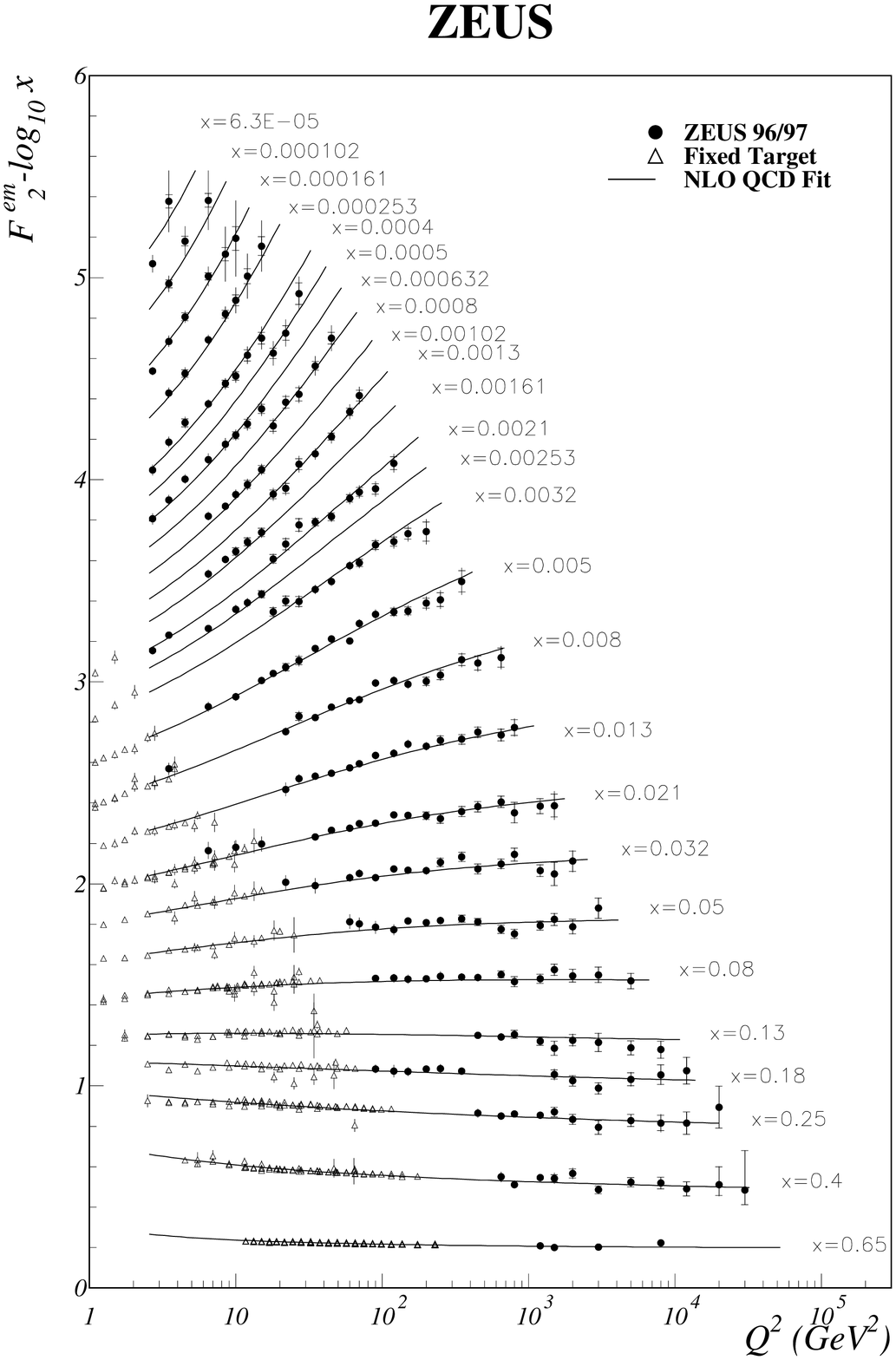}{0}{fig:f2vsq}{
The results for $F_2^{em}$ (points) versus $Q^2$ are shown for fixed $x$.
The fixed target results from NMC, BCDMS and E665
(triangles) and the ZEUS NLO QCD fit (curve)
are also shown.
}

\plot{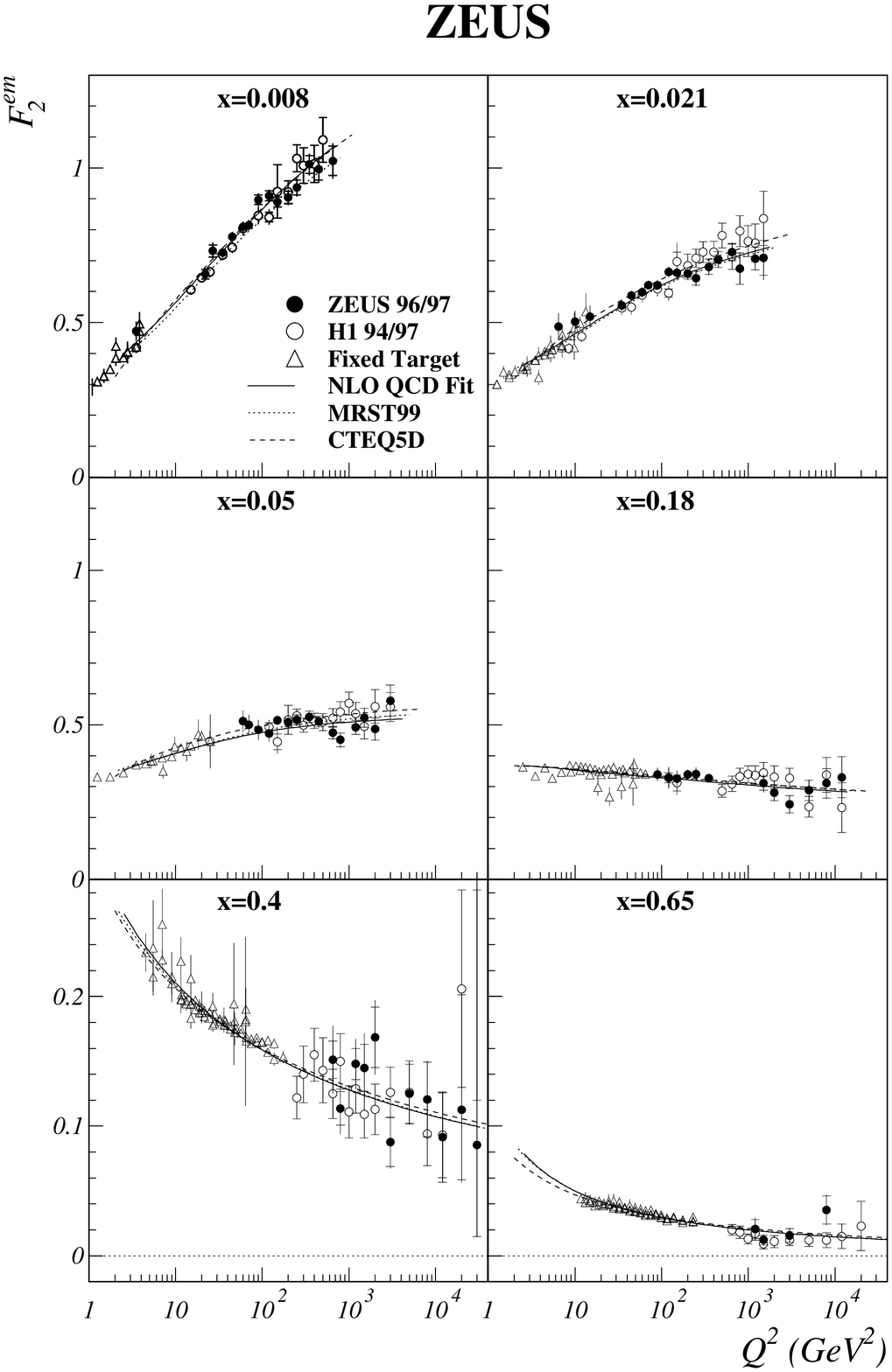}{0}{fig:f2vsq2}{
The results for $F_2^{em}$ versus $Q^2$, for six bins at fixed $x$, 
are compared with results from NMC, BCDMS, E665 (triangles)
and the recently published H1 results (open symbols).
The inner error bars (delimited by the horizontal lines)
show the statistical uncertainties; the outer ones show the statistical and
systematic uncertainties added in quadrature, both for ZEUS and H1.
}

%
%
\end{document}